\documentclass{pasj00}
\draft

\begin{document}
\SetRunningHead{Y.Toba \etal\ }{The 9 and 18 Micron Luminosity Function of Various Types of Galaxies with AKARI}
\Received{2012/08/22}
\Accepted{2013/07/20}
\Published{}

\title{The 9 and 18 Micron Luminosity Function of Various Types of Galaxies with AKARI: Implication for the Dust Torus Structure of AGN}



%
 \author{
   Yoshiki \textsc{Toba}\altaffilmark{1,2},
   Shinki \textsc{Oyabu}\altaffilmark{3},
   Hideo \textsc{Matsuhara}\altaffilmark{2},
   Matthew A \textsc{Malkan}\altaffilmark{4}, 
   Daisuke \textsc{Ishihara}\altaffilmark{3},
   Takehiko \textsc{Wada}\altaffilmark{2},
   Youichi \textsc{Ohyama}\altaffilmark{5},
   Satoshi \textsc{Takita}\altaffilmark{2},
   and
   Chisato \textsc{Yamauchi}\altaffilmark{6}}

 \altaffiltext{1}{Department of Space and Astronautical Science, the Graduate University for Advanced Studies (Sokendai), 3-1-1 Yoshinodai, Chuo-ku, Sagamihara, Kanagawa 252-5210}
 \email{toba@ir.isas.jaxa.jp}
 \altaffiltext{2}{Institute of Space and Astronautical Science, Japan Aerospace Exploration Agency, 3-1-1 Yoshinodai, Chuo-ku, Sagamihara, Kanagawa 252-5210}
 \altaffiltext{3}{Graduate School of Science, Nagoya University, Furo-cho, Chikusa-ku, Nagoya, Aichi 464-8602}
 \altaffiltext{4}{Department of Physics and Astronomy, University of California, Los Angeles, CA 90095-1547, USA}
 \altaffiltext{5}{Institute of Astronomy and Astrophysics, Academia Sinica, P.O. Box 23-141, Taipei 10617, Taiwan, R.O.C} 
 \altaffiltext{6}{Astronomy Data Center, National Astronomical Observatory of Japan, 2-21-1 Osawa, Mitaka, Tokyo 181-8588}  

\KeyWords{galaxies: active --- galaxies: luminosity function, mass function --- galaxies: nuclei --- infrared: galaxies} 

\maketitle

\begin{abstract}
We present the 9 and 18 $\micron$ luminosity functions (LFs) of galaxies at 0.006 $\leq$ z $\leq$ 0.8 (with an average redshift of $\sim$ 0.04) using the AKARI mid-infrared all-sky survey catalog.
 We selected 243 galaxies at 9 $\micron$ and 255 galaxies at 18 $\micron$ from the Sloan Digital Sky Survey (SDSS) spectroscopy region.
 These galaxies were then classified by their optical emission lines, such as the line width of H$\alpha$ or by their emission line ratios of [OIII]/H$\beta$ and [NII]/H$\alpha$ into five types: Type 1 active galactic nuclei (AGN) (Type 1); Type 2 AGN (Type 2); low-ionization narrow emission line galaxies (LINER); galaxies with both star formation and narrow-line AGN activity (composite galaxies); and star-forming galaxies (SF).
 We found that (i) the number density ratio of Type 2 to Type 1 AGNs is 1.73 $\pm$ 0.36, which is larger than a result obtained from the optical LF and (ii) this ratio decreases with increasing 18 $\micron$ luminosity. 
\end{abstract}

\section{Introduction}
The unified model for active galactic nuclei (AGN) postulates that Type 1 and Type 2 AGNs are intrinsically identical.
 The most popular explanation for their observed differences is due to different viewing orientations (\cite{Antonucci}, \cite{Urry}).
 A geometrically thick dusty torus is supposed to surround the AGN central engine.
 When an observer views the central region near the polar axis of the torus, a broad emission line region (BLR) can be observed directly, identifying these galaxies as Type 1 AGNs.
 On the other hand, at angles near the equatorial plane of the torus, the central region of the AGN, with its BLR, cannot be observed. 
 The narrow emission line region (NLR), the size of which is at least an order of magnitude larger than the torus, is always observed regardless of torus orientation.
 Thus, these galaxies are classified as Type 2 AGNs.

The key parameters of dust torus unification-the intrinsic geometry (e.g., thickness and covering factor) and physical properties-are still unknown.
 To investigate these parameters, infrared (IR) observations of AGNs are important, because the reprocessed radiation from the dust in the torus is re-emitted in the IR wavelength range.
 In this study, we focus on the geometrical covering fraction of the dust torus.
 The covering factor (CF) is defined as the fraction of the sky, as seen from the AGN center, which is blocked by obscuring material.
 This corresponds to the fraction of Type 2 AGNs out of the entire AGN population.
 Many previous studies have attempted to estimate CF.
 For instance, \citet{Mor} estimated CF for 26 luminous quasars.
 They fitted the $\sim$2-35 $\micron$ spectra from the Spitzer Space Telescope \citep{Werner} IRS using  three-component models made of a clumpy torus, dusty NLR clouds, and very hot dust clouds.
 The clumpy torus model attempts to fix problems with torus unification. 
 In this more complicated model, the observed differences between Type 1 and Type 2 AGNs are due not only to orientation, but also to random probability (\cite{Nenkovaa}, \cite{Nenkovab}). 
 The observer's line-of-sight passing through the torus may, with some probability, pass through a gap in the clouds. 
 Such an object is classified as a Type 1 AGN, even though it is viewed at a high inclination angle, similar to other Type 2 AGNs.
 Conversely, even when viewed from a near-polar angle, there is some small probability that a dust cloud may happen to obscure the AGN along the particular sightline.
 Such an object is classified as Type 2 AGN.  
 This complex model in general lacks the simplicity and predictive power of the original dusty torus.
 Those authors found a mean CF value of 0.27 by fitting the AGN data, which may be anti-correlated with bolometric luminosity. 
 \citet{Alonso-Herrero+11} also calculated the CF of 13 nearby Seyfert galaxies by using clumpy torus models with a Bayesian approach to fit their IR spectral energy distributions (SEDs) and ground-based high-angular-resolution mid-IR spectroscopy.
 They found a tendency for the CF to be lower ($\sim$0.1-0.3) at high AGN luminosities than at low AGN luminosities ($\sim$0.9-1), which supports the result of \citet{Mor}.
 
However, these studies were limited to small numbers of individual objects, rather than a statistically complete AGN sample. 
 To overcome this problem, we estimate the CF of dust torus using the IR luminosity function (LF). 
 The LF of galaxies is a fundamental statistical tool for describing galaxy properties, since it should be almost entirely independent of the viewing angle. 
 In this study, we construct the LF using the AKARI satellite. 
 AKARI, the first Japanese space satellite dedicated to infrared astronomy, was launched in 2006 \citep{Murakami}.
 One of its most important results was an all-sky survey in the mid-IR (MIR) and far-IR (FIR) (\cite{Ishihara}, \cite{Yamamura}).
 The spatial resolution and sensitivity of AKARI are much better than those of the Infrared Astronomical Satellite (IRAS: \cite{Neugebauer}, \cite{Beichman}) which performed a previous all-sky IR survey.
 In particular, the detection limits (5$\sigma$) for point sources per scan are 50 and 90 mJy for the 9 and 18 $\micron$ bands, respectively, with spatial resolutions of about \timeform{5''}, thus surpassing the IRAS survey in its 12 and 25 $\micron$ bands by an order of magnitude in both sensitivity and spatial resolution.

In this study, we report the 9 and 18 $\micron$ LFs of galaxies at 0.006 $\leq z \leq$ 0.8 using the AKARI MIR all-sky survey catalog.
 The sample selection is described in section 2.
 In section 3, we present the 9 and 18 $\micron$ LFs with the 1/$V_{\mathrm{max}}$ technique and compare our results with previous studies.
 In section 4, we discuss the number density ratio of Type 1 AGNs and Type 2 AGNs as estimated from the LFs of AGNs and compare it to previous optical and X-ray estimates.
 This study provides us with an important estimate of the dust torus structure of AGNs.
 Throughout this study, we assume a flat universe with $\Omega_k =0$, and we adopt ($\Omega_M$, $\Omega_{\Lambda}$)=(0.3, 0.7) and $H_0$=75 km s$^{-1}$ Mpc$^{-1}$.

\section{Data and Analysis}
\subsection{Sample Selection}
\label{sample_selection}

\begin{figure}
   \begin{center}
      \FigureFile(80mm,50mm){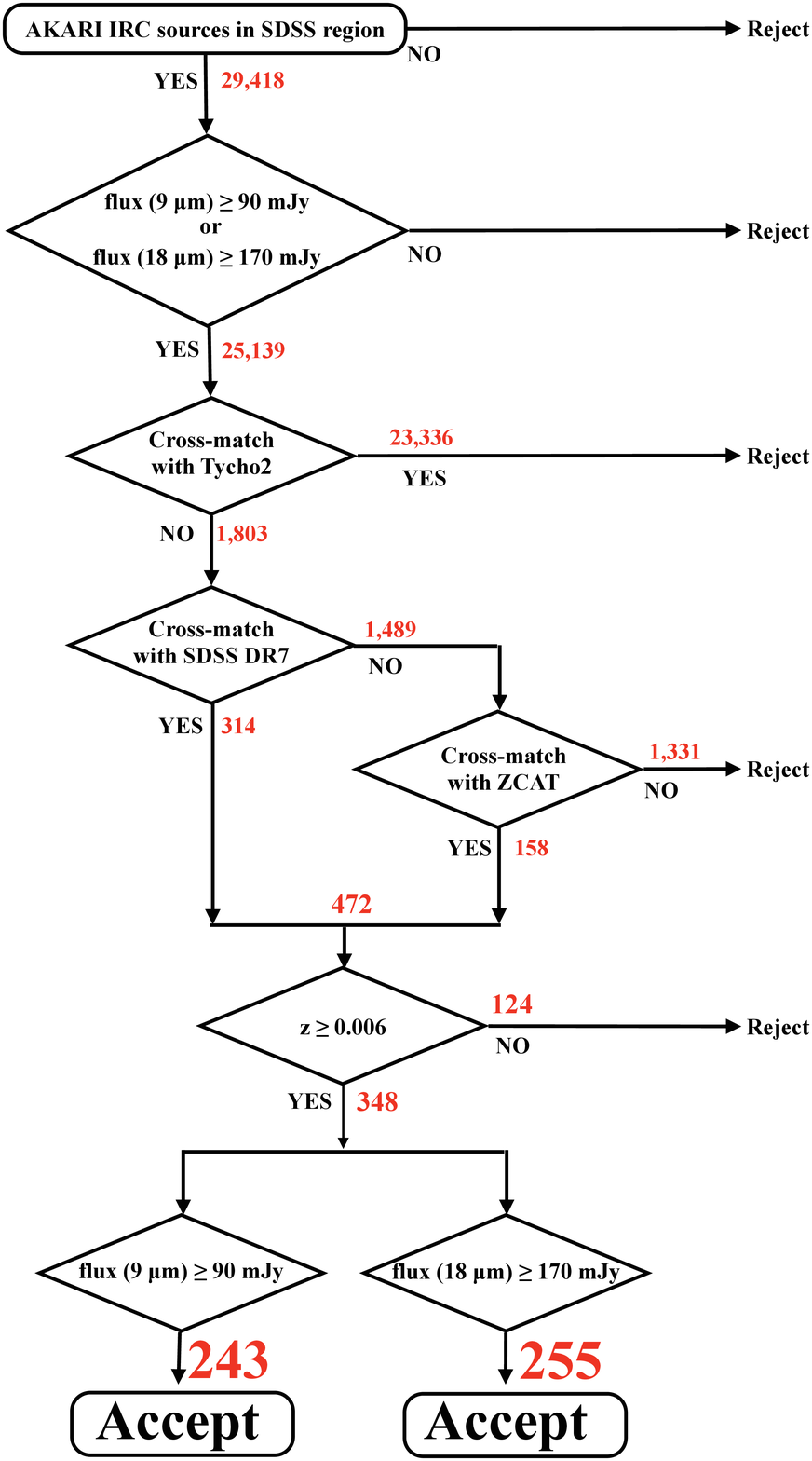}
   \end{center}
\caption{Flow chart for sample selection process.}
\label{sample-selection-algorithm}
\end{figure}

\begin{figure*}
   \begin{center}
      \FigureFile(160mm,50mm){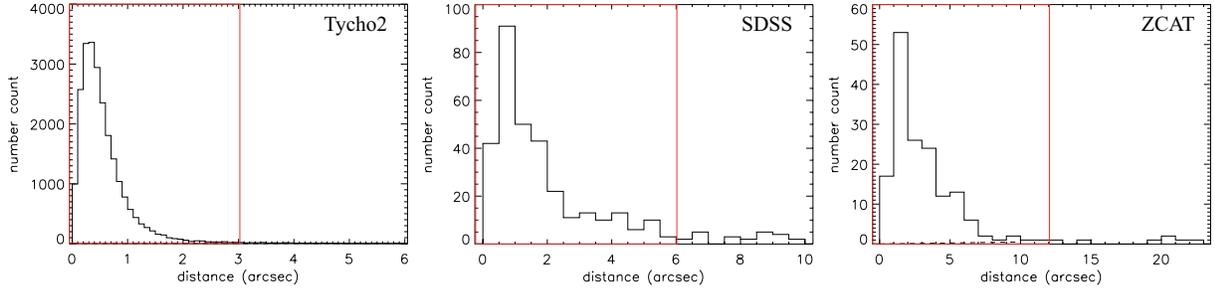}
   \end{center}
\caption{Histogram of the angular separation of AKARI sources from the Tycho 2 (left), SDSS (middle), and ZCAT (right) coordinates. We adopt search radii for the Tycho 2, SDSS, and ZCAT as 3, 6, and 12 arcsec, respectively. The red frame represents the search radii used.}
\label{cross-match_NED_SDSS}
\end{figure*}

The AKARI MIR all-sky survey catalog provides the positions and fluxes of 870,973 sources (including 844,649 sources in the 9 $\micron$ band and 194,551 sources in the 18 $\micron$ band).
 The sample used for this study was selected from AKARI MIR sources with spectroscopy from the Sloan Digital Sky Survey (SDSS) Data Release 7 \citep{Abazajian} and the Center for Astrophysics (CfA) redshift survey (ZCAT: \cite{Huchra}).  
 The flow chart used for sample selection is shown in figure \ref{sample-selection-algorithm}.
 First, our sample was narrowed to the 29,418 AKARI sources in the SDSS spectroscopic region.
 To ensure flux accuracy, we then extracted objects that met the following criteria:
\begin{itemize}
\item flux (9   $\micron$)   $\geq$   90 mJy (for  9 $\micron$ LF) 
\item flux (18  $\micron$)   $\geq$  170 mJy (for 18 $\micron$ LF)
\end{itemize}
These fluxes correspond to the 50\% completeness flux limits calculated by \citet{Kataza}.
 Next the 25,139 remaining objects (24,936 sources with 9 $\micron$ flux greater than 90 mJy and 4,316 sources with 18 $\micron$ flux greater than 170 mJy) were cross-matched with the Tycho-2 Catalog (H\o g \etal\ 2000) to remove the galactic bright stars.
 This catalog contains the positions, proper motions, and two-color photometry for the 2.5 million bright stars in the sky down to the magnitude limit of the plates ($V_T \sim11.5$).
 To avoid the omission of high proper-motion stars, we referred to the \textit{mean position}, rigorously propagated to the epoch J2000.0, by the proper motions in this catalog. 
 As a result, a total of 23,336 ($\sim$92.8\% of the objects cross-identified with the Tycho-2 catalog) stars (hereinafter AKARI-Tycho 2 objects) were extracted.  
 As shown in figure \ref{cross-match_NED_SDSS}, we adopted 3 arcsec as our search radius, because the star density in the SDSS spectroscopic region is at most $\sim$50 deg$^{-1}$ (H\o g \etal\ 2000).
 Thus, by adopting this search radius, the probability of chance coincidence is less than 0.01\% (i.e., 25,139 $\times$ 0.0001 $\sim$ 3 sources may be misidentified), which is acceptable.
 
\begin{figure}
   \begin{center}
      \FigureFile(80mm,50mm){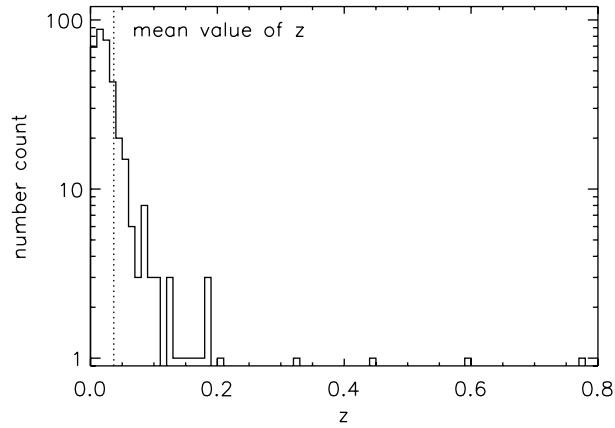}
   \end{center}
\caption{Redshift distribution of the 348 selected galaxies. The mean value of redshift is represented by the dotted line.}
\label{z_dist}
\end{figure}

We then identified 1,803 AKARI sources in the SDSS spectroscopic catalog and in the ZCAT.
 Both have the redshifts of $\sim$900,000 objects cataloged. 
 The SDSS spectroscopic catalog mainly contains magnitude-limited objects with \citet{Petrosian+76} magnitude brighter than $r$ = 17.77 for galaxy samples \citep{Strauss} and with point-spread
function (PSF) magnitude brighter than $i$ = 19.1 for quasar samples at $z \leq$ 2.3 \citep{Richards}.  
 The advantage of using the SDSS is that we obtain uniform spectroscopic data.
 The average redshift of these spectroscopic objects is $\sim$0.1.
 However, bright objects that might be saturated are excluded in the spectroscopic target selection (\cite{Eisenstein}, \cite{Strauss}).
 Therefore, for bright galaxies, we use the ZCAT, which is compiled from literature, and has a limiting magnitude of $\sim$15.5 in the B band.
 The search radii for the SDSS and ZCAT were 6 and 12 arcsec, respectively, as shown in figure \ref{cross-match_NED_SDSS}.
 For the SDSS, the surface density of galaxies in the SDSS spectroscopic region is 92 deg$^{-2}$ \citep{Strauss}.
  Within 6 arcsec around each AKARI MIR source, the expected number of galaxies is then 0.008.
  Adopting this search radius means that the probability of chance coincidence is less than 0.1\% (i.e., 1,803 $\times$ 0.001 $\sim$ 2 sources may be misidentified), which is acceptable. 
 For the ZCAT, the surface density of galaxies in the SDSS spectroscopic region is $\sim$71 deg$^{-2}$.
 Thus, by adopting a search radius of 12 arcsec, the probability of chance coincidence is less than 0.25\% for the AKARI-ZCAT sources (i.e., 1,489 $\times$ 0.0025 $\sim$ 4 sources may be misidentified). 
 This value is relatively larger than the corresponding value of the SDSS, but is acceptable, because the ZCAT assembles the nearby or spatially extended galaxies. 
 At this point, 472 galaxies remained for selection.
 Furthermore, we excluded 124 local galaxies with $z \leq$ 0.006 from the sample.
 The errors in distance measurement are dominated by peculiar motions for galaxies with $z \leq$ 0.006, and thus, the luminosity also has a large error.
 A final sample of 348 galaxies was selected.
 The mean value of their redshifts is 0.04, and the redshift distribution is shown in figure \ref{z_dist}.
 The details of the 348 galaxies selected are given in table \ref{sample_list}. 
 Ultimately, 243 and 255 galaxies were selected for construction of the 9 $\micron$ LF and the 18 $\micron$ LF, respectively.
 Throughout this process, we mainly used the AKARI Catalog Archive Server \citep{Yamauchi_a} and a software kit called ``2MASS Catalog Server Kit'' to easily construct a high-performance database server for the Two Micron All Sky Survey (2MASS) Point Source Catalog (which includes 470,992,970 objects) and several all-sky catalogs \citep{Yamauchi_b}.

 \begin{small}
\begin{longtable}{lrrrrrcc}
\caption{List of 348 galaxies.}
\label{sample_list}
\hline 
\hline
\multicolumn{1}{c}{name}	&	\multicolumn{1}{c}{RA}	&	\multicolumn{1}{c}{DEC}	&	\multicolumn{1}{c}{$f_{9\mu m}$}	&	\multicolumn{1}{c}{$f_{18\mu m}$}	&	\multicolumn{1}{c}{z}	&	\multicolumn{1}{c}{type} & \multicolumn{1}{c}{ref}\\
		&	\multicolumn{1}{c}{j2000}	&	\multicolumn{1}{c}{j2000}	&	\multicolumn{1}{c}{(Jy)}	&	\multicolumn{1}{c}{(Jy)}	&	&  & \\	
\hline
\endhead
\hline
\endfoot
\hline
\multicolumn{8}{l}{\hbox to 10pt{\parbox{160mm}{\footnotesize
(1) SDSS; (2) SIMBD; (3) NED; (4) \citet{Kewley+06}; (5) \citet{Kirhakos}; (6) \citet{Clavel}; (7) \citet{Parra}; (8) \citet{Corbett}; (9) \citet{Burlon}; (10) \citet{Wang}; (11) \citet{Ho}; (12) \citet{Veilleux+95}; (13) \citet{Cohen}; (14) \citet{Bukhmastova}; (15) \citet{Konstantopoulos}; (16) \citet{Koulouridis}; (17) \citet{Yuan}; (18) \citet{Leech}; (19) \citet{Wu}; (20) \citet{Petrosian+07}; (21) \citet{Petrosian+08}; (22) \citet{Moustakas}; (23) \citet{Veron-Cetty}; (24) \citet{Alonso-Herrero}; (25) \citet{Poggianti}; (26) Okayama obs.; (27) \citet{Caccianiga}
 }}}
 \endlastfoot
2MASX J00090793+1427558 & 00:09:07.90 & +14:27:56.7 & 0.052 & 0.248 & 0.041 & Type 2     & (1)\\
ARP 256 (3)01 & 00:18:50.92 & -10:22:37.4 & 0.178 & 0.517 & 0.027 & Composite    & (1)\\
NGC 0192 & 00:39:13.39 & +00:51:49.4 & 0.185 & \multicolumn{1}{c}{---} & 0.014 & SF      & (4)\\
NGC 0291 & 00:53:29.91 & -08:46:02.9 & \multicolumn{1}{c}{---} & 0.204 & 0.019 & Type 2     & (1)\\
UGC 00774 & 01:13:50.91 & +13:16:19.2 & 0.123 & 0.324 & 0.050 & Type 1     & (2),(3)\\
CGCG 436-030 & 01:20:02.61 & +14:21:42.4 & 0.182 & 0.592 & 0.031 & Composite    & (1)\\
2MASX J01221811+0100262 & 01:22:17.93 & +01:00:26.1 & \multicolumn{1}{c}{---} & 0.306 & 0.055 & SF      & (1)\\
MRK 0995 & 01:27:29.21 & -08:33:14.0 & 0.132 & 0.150 & 0.049 & Composite    & (1)\\
GIN 086 & 01:37:06.98 & -09:08:57.7 & \multicolumn{1}{c}{---} & 0.197 & 0.070 & Type 2     & (1)\\
UGC 01260 & 01:48:33.18 & +12:36:50.1 & 0.141 & 0.214 & 0.018 & SF      & (3)\\
SDSS J015103.75-094209.8 & 01:51:03.66 & -09:42:12.1 & 0.259 & 0.348 & 0.006 & SF      & (6)\\
NGC 0716 & 01:52:59.63 & +12:42:30.8 & 0.262 & 0.342 & 0.015 & LINER   & (7)\\
MRK 1014 & 01:59:50.31 & +00:23:39.7 & \multicolumn{1}{c}{---} & 0.244 & 0.163 & Type 1     & (1)\\
NGC 0835 & 02:09:24.61 & -10:08:09.5 & 0.247 & 0.342 & 0.014 & Type 2     & (2),(3)\\
IC 0210 & 02:09:28.24 & -09:40:46.9 & 0.185 & \multicolumn{1}{c}{---} & 0.006 & Unknown & \multicolumn{1}{c}{---}\\
NGC 0838 & 02:09:38.53 & -10:08:47.0 & 0.514 & 0.861 & 0.013 & SF      & (1)\\
NGC 0839 & 02:09:42.76 & -10:11:02.5 & 0.297 & 1.101 & 0.013 & SF      & (8)\\
NGC 0863 & 02:14:33.53 & -00:46:00.8 & \multicolumn{1}{c}{---} & 0.266 & 0.026 & Type 1     & (2),(3)\\
NGC 0905 & 02:22:43.69 & -08:43:10.6 & \multicolumn{1}{c}{---} & 0.188 & 0.046 & Unknown & \multicolumn{1}{c}{---}\\
MRK 1044 & 02:30:05.57 & -08:59:53.4 & 0.110 & \multicolumn{1}{c}{---} & 0.016 & Type 1     & (3)\\
UGC 02024 & 02:33:01.26 & +00:25:15.2 & 0.160 & 0.456 & 0.022 & Type 2     & (1)\\
SDSS J023437.83-084716.0 & 02:34:37.77 & -08:47:16.7 & 0.165 & 0.368 & 0.043 & Type 1     & (2)\\
NGC 1142 & 02:55:12.06 & -00:11:02.8 & 0.265 & 0.380 & 0.029 & Type 2     & (2),(3)\\
UGC 02403 & 02:55:57.28 & +00:41:31.4 & 0.205 & 0.392 & 0.014 & SF      & (1)\\
NGC 1194 & 03:03:49.08 & -01:06:12.6 & 0.169 & 0.415 & 0.014 & Type 2     & (1)\\
MRK 0609 & 03:25:25.38 & -06:08:38.8 & 0.125 & \multicolumn{1}{c}{---} & 0.034 & Type 2     & (1)\\
2MASX J03474022+0105143 & 03:47:40.15 & +01:05:13.8 & 0.166 & 0.407 & 0.031 & Type 1     & (2),(3)\\
UGC 03973 & 07:42:32.81 & +49:48:35.1 & 0.276 & 0.611 & 0.022 & Type 1     & (1)\\
MCG +05-19-001 NED02 & 07:44:09.11 & +29:14:51.0 & 0.160 & 0.275 & 0.016 & Type 2     & (2)\\
NGC 2445 $[ASR92]$ A & 07:46:55.02 & +39:00:54.5 & 0.102 & 0.223 & 0.013 & SF      & (3)\\
UGC 04132 & 07:59:12.92 & +32:54:49.5 & 0.242 & \multicolumn{1}{c}{---} & 0.017 & SF      & (1)\\
NGC 2498 & 07:59:38.76 & +24:58:56.6 & \multicolumn{1}{c}{---} & 0.346 & 0.016 & Unknown & \multicolumn{1}{c}{---}\\
UGC 04145 & 07:59:40.13 & +15:23:12.4 & 0.174 & 0.577 & 0.016 & Type 2     & (1)\\
SBS 0755+509 & 07:59:40.87 & +50:50:24.3 & \multicolumn{1}{c}{---} & 0.178 & 0.055 & Type 2     & (1)\\
CGCG 118-036 & 07:59:53.45 & +23:23:23.8 & 0.105 & \multicolumn{1}{c}{---} & 0.029 & Type 2     & (1)\\
SBS 0756+553 & 08:00:33.52 & +55:13:00.2 & \multicolumn{1}{c}{---} & 0.217 & 0.035 & Composite    & (1)\\
2MASX J08025293+2552551 & 08:02:52.96 & +25:52:55.0 & \multicolumn{1}{c}{---} & 0.326 & 0.081 & Type 2     & (1)\\
NGC 2512 & 08:03:08.01 & +23:23:31.2 & 0.156 & 0.361 & 0.016 & SF      & (3)\\
MRK 1212 & 08:07:05.48 & +27:07:33.4 & \multicolumn{1}{c}{---} & 0.213 & 0.041 & Composite    & (1)\\
IC 2227 & 08:07:07.14 & +36:14:00.4 & 0.228 & 0.224 & 0.032 & Type 2     & (1)\\
UGC 04229 & 08:07:41.07 & +39:00:14.8 & \multicolumn{1}{c}{---} & 0.209 & 0.023 & Type 2     & (1)\\
2MASX J08100697+1838176 & 08:10:07.00 & +18:38:17.5 & \multicolumn{1}{c}{---} & 0.418 & 0.016 & Composite    & (1)\\
NGC 2538 & 08:11:23.15 & +03:37:59.0 & 0.104 & \multicolumn{1}{c}{---} & 0.013 & SF      & (4)\\
CGCG 031-072 & 08:14:25.33 & +04:20:32.7 & 0.131 & \multicolumn{1}{c}{---} & 0.033 & Type 1     & (9)\\
UGC 04306 & 08:17:36.68 & +35:26:44.9 & 0.178 & 0.240 & 0.008 & SF      & (1)\\
NGC 2561 & 08:19:36.98 & +04:39:26.5 & 0.140 & \multicolumn{1}{c}{---} & 0.014 & Unknown & \multicolumn{1}{c}{---}\\
2MASX J08244333+2959238 & 08:24:43.31 & +29:59:23.3 & 0.121 & \multicolumn{1}{c}{---} & 0.025 & Type 2     & (1)\\
MRK 0091 & 08:32:27.96 & +52:36:21.8 & 0.143 & 0.307 & 0.017 & SF      & (1)\\
NGC 2623 & 08:38:24.05 & +25:45:16.7 & 0.177 & 0.593 & 0.018 & Unknown & \multicolumn{1}{c}{---}\\
FBQS J084215.2+402533 & 08:42:15.32 & +40:25:32.7 & 0.138 & 0.272 & 0.055 & Type 2     & (1)\\
KUG 0842+400 & 08:46:03.97 & +39:49:48.7 & 0.126 & \multicolumn{1}{c}{---} & 0.029 & SF      & (1)\\
IC 2406 & 08:48:04.58 & +17:42:10.7 & 0.149 & \multicolumn{1}{c}{---} & 0.016 & Composite    & (1)\\
OJ +287:$[YJI97]$ EXT & 08:54:48.83 & +20:06:30.7 & \multicolumn{1}{c}{---} & 0.206 & 0.306 & Type 1     & (3)\\
NGC 2718 & 08:58:50.32 & +06:17:35.6 & 0.146 & 0.300 & 0.013 & SF      & (1)\\
2MASX J09002536+3903542 & 09:00:25.37 & +39:03:53.8 & 0.312 & 0.794 & 0.058 & Composite    & (1)\\
UGC 04730 & 09:01:58.45 & +60:09:05.9 & 0.106 & \multicolumn{1}{c}{---} & 0.011 & Composite    & (1)\\
NGC 2731 & 09:02:08.40 & +08:18:04.8 & 0.181 & \multicolumn{1}{c}{---} & 0.009 & SF      & (10)\\
NGC 2738 & 09:04:00.38 & +21:58:02.2 & 0.198 & \multicolumn{1}{c}{---} & 0.010 & SF      & (1)\\
NGC 2750 & 09:05:47.91 & +25:26:14.0 & 0.121 & 0.197 & 0.009 & SF      & (11)\\
NGC 2761 & 09:07:30.78 & +18:26:04.6 & 0.194 & 0.280 & 0.029 & Composite    & (1)\\
NGC 2764 & 09:08:17.47 & +21:26:34.6 & 0.228 & \multicolumn{1}{c}{---} & 0.009 & Unknown & \multicolumn{1}{c}{---}\\
NGC 2773 & 09:09:44.23 & +07:10:25.2 & 0.187 & 0.182 & 0.018 & SF      & (1)\\
FIRST J091345.4+405628 & 09:13:45.41 & +40:56:27.9 & \multicolumn{1}{c}{---} & 0.205 & 0.442 & Type 2     & (2),(3)\\
NGC 2782:$[HK83]$ 04 & 09:14:05.09 & +40:06:49.2 & 0.314 & 0.692 & 0.009 & SF      & (11)\\
NGC 2789 & 09:14:59.67 & +29:43:47.9 & 0.125 & \multicolumn{1}{c}{---} & 0.021 & Unknown & \multicolumn{1}{c}{---}\\
NGC 2785 & 09:15:15.38 & +40:55:03.0 & 0.377 & 0.604 & 0.009 & Type 2     & (1)\\
UGC 04881 & 09:15:55.48 & +44:19:53.2 & \multicolumn{1}{c}{---} & 0.182 & 0.040 & LINER   & (1)\\
SDSS J091722.85+415959.0 & 09:17:22.87 & +41:59:59.3 & 0.572 & 1.380 & 0.006 & Composite    & (1)\\
MRK 0704 & 09:18:25.94 & +16:18:18.8 & 0.256 & 0.469 & 0.029 & Type 1     & (2),(3)\\
CGCG 121-075 & 09:23:42.91 & +22:54:30.8 & 0.078 & 0.178 & 0.033 & Type 1     & (1)\\
NGC 2854 & 09:24:03.16 & +49:12:15.5 & 0.186 & \multicolumn{1}{c}{---} & 0.009 & Unknown & \multicolumn{1}{c}{---}\\
NGC 2856 & 09:24:16.08 & +49:14:57.2 & 0.239 & 0.466 & 0.009 & Unknown & \multicolumn{1}{c}{---}\\
UGC 05025 & 09:26:03.31 & +12:44:04.0 & 0.099 & 0.214 & 0.029 & Type 1     & (1)\\
UGC 05046 & 09:28:06.66 & +17:11:47.2 & 0.119 & \multicolumn{1}{c}{---} & 0.014 & SF      & (1)\\
CGCG 181-049 & 09:30:11.21 & +34:39:53.0 & 0.096 & \multicolumn{1}{c}{---} & 0.023 & SF      & (1)\\
SBS 0927+493 & 09:31:06.69 & +49:04:47.7 & \multicolumn{1}{c}{---} & 0.228 & 0.034 & LINER   & (1)\\
UGC 05101 & 09:35:51.63 & +61:21:13.4 & 0.135 & 0.457 & 0.039 & Type 2     & (1)\\
CGCG 239-011 NED02 & 09:36:37.14 & +48:28:28.1 & 0.147 & 0.330 & 0.026 & SF      & (12)\\
CGCG 122-055 & 09:42:04.77 & +23:41:07.2 & \multicolumn{1}{c}{---} & 0.231 & 0.021 & Type 1     & (1)\\
NGC 2966 & 09:42:11.49 & +04:40:23.2 & 0.178 & 0.354 & 0.007 & SF      & (4)\\
2MASX J09452133+1737533 & 09:45:21.29 & +17:37:52.8 & \multicolumn{1}{c}{---} & 0.286 & 0.128 & Type 2     & (1)\\
NGC 2990 & 09:46:17.13 & +05:42:31.8 & 0.151 & \multicolumn{1}{c}{---} & 0.010 & SF      & (1)\\
IC 0563 & 09:46:20.29 & +03:02:46.4 & 0.177 & \multicolumn{1}{c}{---} & 0.020 & SF      & (1)\\
I 564 & 09:46:20.87 & +03:04:13.0 & 0.183 & \multicolumn{1}{c}{---} & 0.020 & Composite    & (1)\\
CGCG 007-035 & 09:47:13.53 & +00:39:54.9 & 0.097 & \multicolumn{1}{c}{---} & 0.020 & Composite    & (1)\\
NGC 3055 & 09:55:17.98 & +04:16:11.5 & 0.156 & 0.304 & 0.006 & Composite    & (1)\\
IC 2519 & 09:55:58.85 & +34:02:10.9 & 0.046 & 0.175 & 0.021 & SF      & (1)\\
UGC 05376 & 10:00:27.03 & +03:22:26.4 & 0.279 & \multicolumn{1}{c}{---} & 0.007 & SF      & (1)\\
NGC 3094 & 10:01:25.92 & +15:46:11.6 & 0.707 & 1.505 & 0.008 & LINER   & (13)\\
SDSS J100131.21+465946.7 & 10:01:31.28 & +46:59:46.8 & \multicolumn{1}{c}{---} & 0.221 & 0.086 & Type 2     & (1)\\
3C 234 & 10:01:49.52 & +28:47:08.6 & 0.112 & 0.274 & 0.185 & Type 2     & (1)\\
UGC 05403 & 10:02:35.57 & +19:10:37.6 & 0.140 & 0.266 & 0.007 & SF      & (1)\\
CGCG 064-055 & 10:05:51.15 & +12:57:40.2 & \multicolumn{1}{c}{---} & 0.291 & 0.009 & Type 2     & (1)\\
IC 2551 & 10:10:40.30 & +24:24:50.0 & 0.145 & 0.443 & 0.021 & Composite    & (1)\\
2MASX J10104337+0612013 & 10:10:43.30 & +06:12:01.4 & \multicolumn{1}{c}{---} & 0.213 & 0.098 & Type 1     & (1)\\
NGC 3154 & 10:13:01.20 & +17:02:03.4 & 0.254 & 0.393 & 0.022 & SF      & (1)\\
UGC 05613 & 10:23:32.60 & +52:20:31.4 & 0.183 & \multicolumn{1}{c}{---} & 0.032 & Composite    & (1)\\
WAS 11 & 10:27:25.87 & +20:26:50.7 & 0.130 & \multicolumn{1}{c}{---} & 0.019 & Composite    & (1)\\
MRK 0034 & 10:34:08.52 & +60:01:51.4 & \multicolumn{1}{c}{---} & 0.291 & 0.051 & Type 2     & (1)\\
SDSS J103631.87+022144.0 & 10:36:31.91 & +02:21:44.2 & 0.124 & 0.311 & 0.050 & Composite    & (1)\\
NGC 3306 & 10:37:10.23 & +12:39:07.8 & 0.154 & \multicolumn{1}{c}{---} & 0.010 & SF      & (10)\\
IC 2598 & 10:39:42.34 & +26:43:38.3 & 0.144 & 0.219 & 0.019 & SF      & (1)\\
MRK 0726 & 10:45:49.87 & +27:37:12.7 & \multicolumn{1}{c}{---} & 0.186 & 0.044 & Composite    & (1)\\
NGC 3367 & 10:46:34.89 & +13:45:00.8 & 0.168 & 0.606 & 0.010 & Composite    & (1)\\
MRK 0727 & 10:48:44.16 & +26:03:13.4 & \multicolumn{1}{c}{---} & 0.240 & 0.026 & SF      & (1)\\
UGC 05941 NED02 & 10:50:21.58 & +41:27:51.0 & 0.094 & 0.113 & 0.024 & SF      & (1)\\
2MASX J10563881+1419306 & 10:56:38.83 & +14:19:28.9 & \multicolumn{1}{c}{---} & 0.209 & 0.081 & Type 1     & (1)\\
NGC 3471 & 10:59:09.07 & +61:31:51.9 & 0.272 & 0.563 & 0.007 & SF      & (1)\\
2MASX J10591815+2432343 & 10:59:18.11 & +24:32:34.3 & 0.131 & 0.452 & 0.043 & LINER   & (2)\\
UGC 06074 & 10:59:58.26 & +50:54:10.6 & \multicolumn{1}{c}{---} & 0.421 & 0.010 & Composite    & (1)\\
2MASX J11001238+0846157 & 11:00:12.38 & +08:46:12.1 & 0.133 & 0.408 & 0.100 & Type 2     & (1)\\
UGC 06100 & 11:01:34.18 & +45:39:16.2 & \multicolumn{1}{c}{---} & 0.181 & 0.029 & Type 2     & (1)\\
UGC 06103 & 11:01:59.13 & +45:13:42.1 & 0.133 & \multicolumn{1}{c}{---} & 0.020 & Composite    & (1)\\
CGCG 241-078 & 11:06:37.46 & +46:02:19.2 & \multicolumn{1}{c}{---} & 0.193 & 0.025 & Composite    & (1)\\
NGC 3561A & 11:11:12.95 & +28:42:42.7 & 0.127 & 0.285 & 0.029 & Composite    & (1)\\
IC 2637 & 11:13:49.77 & +09:35:10.7 & 0.149 & 0.155 & 0.029 & Type 2     & (1)\\
NGC 3583 & 11:14:10.89 & +48:19:06.7 & 0.428 & \multicolumn{1}{c}{---} & 0.007 & Unknown & \multicolumn{1}{c}{---}\\
B2 1111+32 & 11:14:38.97 & +32:41:33.5 & 0.087 & 0.246 & 0.189 & Type 1     & (2),(3)\\
NGC 3633 & 11:20:26.30 & +03:35:07.4 & 0.193 & 0.339 & 0.009 & Composite    & (1)\\
NGC 3656 & 11:23:38.46 & +53:50:33.5 & 0.110 & \multicolumn{1}{c}{---} & 0.010 & Composite    & (1)\\
FIRST J112443.5+384546 & 11:24:43.62 & +38:45:46.7 & 0.111 & \multicolumn{1}{c}{---} & 0.007 & SF      & (11)\\
IC 2810 & 11:25:44.92 & +14:40:35.2 & \multicolumn{1}{c}{---} & 0.208 & 0.034 & Composite    & (1)\\
IC 2846 & 11:28:00.48 & +11:09:28.3 & \multicolumn{1}{c}{---} & 0.209 & 0.041 & LINER   & (1)\\
ARP 299:$[ZWM2003]$ 04 & 11:28:30.91 & +58:33:42.7 & 1.668 & 7.138 & 0.010 & Composite    & (1)\\
SDSS J112833.61+583346.5 & 11:28:33.55 & +58:33:46.2 & 0.882 & 4.616 & 0.010 & Composite    & (1)\\
IC 0698 & 11:29:03.86 & +09:06:42.3 & 0.171 & 0.166 & 0.021 & Composite    & (1)\\
CGCG 126-075 & 11:31:03.70 & +20:14:08.6 & \multicolumn{1}{c}{---} & 0.192 & 0.014 & Composite    & (1)\\
NGC 3714 & 11:31:53.59 & +28:21:29.3 & 0.128 & \multicolumn{1}{c}{---} & 0.024 & Composite    & (1)\\
UGC 06527 NED03 & 11:32:40.32 & +52:57:02.4 & \multicolumn{1}{c}{---} & 0.210 & 0.027 & Type 2     & (1)\\
CGCG 126-101 & 11:34:50.52 & +25:31:50.4 & 0.171 & \multicolumn{1}{c}{---} & 0.024 & Unknown & \multicolumn{1}{c}{---}\\
SBS 1132+579 & 11:35:24.82 & +57:38:59.5 & 0.099 & \multicolumn{1}{c}{---} & 0.029 & SF      & (1)\\
SBS 1133+572 & 11:35:49.09 & +56:57:08.8 & 0.114 & 0.343 & 0.051 & Type 2     & (1)\\
NGC 3758 & 11:36:29.20 & +21:35:48.1 & 0.119 & 0.190 & 0.030 & Type 1     & (1)\\
CGCG 097-022 & 11:36:39.87 & +17:38:35.3 & \multicolumn{1}{c}{---} & 0.222 & 0.027 & Composite    & (1)\\
NGC 3781 & 11:39:03.80 & +26:21:40.7 & 0.073 & 0.398 & 0.023 & Type 1     & (1)\\
NGC 3800 & 11:40:13.51 & +15:20:32.0 & 0.320 & \multicolumn{1}{c}{---} & 0.011 & Unknown & \multicolumn{1}{c}{---}\\
NGC 3822 & 11:42:11.10 & +10:16:39.7 & 0.225 & \multicolumn{1}{c}{---} & 0.021 & Type 2     & (2),(3)\\
SDSS J114212.21+002004.0 & 11:42:12.31 & +00:20:04.7 & 0.090 & 0.378 & 0.019 & SF      & (1)\\
NGC 3839 & 11:43:54.33 & +10:47:06.0 & 0.183 & 0.218 & 0.020 & SF      & (1)\\
MRK 0428 & 11:44:10.96 & +37:11:12.5 & 0.104 & 0.446 & 0.042 & Composite    & (1)\\
UGC 06732 & 11:45:33.05 & +58:58:40.7 & \multicolumn{1}{c}{---} & 0.190 & 0.010 & SF      & (14)\\
NGC 3849 & 11:45:35.14 & +03:13:53.1 & 0.159 & \multicolumn{1}{c}{---} & 0.020 & SF      & (1)\\
ARP 248 NED02 & 11:46:45.28 & -03:50:51.4 & 0.139 & 0.293 & 0.017 & SF      & (10)\\
NGC 3888 & 11:47:34.15 & +55:58:02.4 & 0.195 & \multicolumn{1}{c}{---} & 0.008 & SF      & (1)\\
IC 0737 & 11:48:27.56 & +12:43:38.5 & \multicolumn{1}{c}{---} & 0.247 & 0.014 & Composite    & (15)\\
NGC 3934 & 11:52:12.58 & +16:51:06.0 & 0.106 & \multicolumn{1}{c}{---} & 0.013 & Unknown & \multicolumn{1}{c}{---}\\
NGC 3935 & 11:52:24.06 & +32:24:14.0 & \multicolumn{1}{c}{---} & 0.306 & 0.010 & LINER   & (1)\\
UGC 06901 & 11:55:38.05 & +43:02:45.6 & 0.163 & \multicolumn{1}{c}{---} & 0.024 & SF      & (1)\\
NGC 3987 & 11:57:20.93 & +25:11:42.9 & 0.216 & \multicolumn{1}{c}{---} & 0.015 & Unknown & \multicolumn{1}{c}{---}\\
NGC 3994 & 11:57:36.89 & +32:16:39.1 & 0.258 & \multicolumn{1}{c}{---} & 0.010 & LINER   & (1)\\
IC 0751 & 11:58:52.49 & +42:34:12.3 & \multicolumn{1}{c}{---} & 0.201 & 0.031 & Type 2     & (1)\\
UGC 07017 & 12:02:22.46 & +29:51:42.2 & 0.216 & 0.213 & 0.010 & SF      & (1)\\
NGC 4045 & 12:02:42.34 & +01:58:37.5 & 0.241 & 0.321 & 0.007 & LINER   & (1)\\
NGC 4047 & 12:02:50.75 & +48:38:11.0 & 0.250 & \multicolumn{1}{c}{---} & 0.011 & Unknown & \multicolumn{1}{c}{---}\\
UGC 07064 & 12:04:43.30 & +31:10:35.8 & \multicolumn{1}{c}{---} & 0.186 & 0.025 & Type 2     & (1)\\
CGCG 098-059 & 12:07:09.55 & +16:59:43.7 & 0.146 & 0.284 & 0.022 & SF      & (1)\\
NGC 4152 & 12:10:37.46 & +16:01:59.8 & \multicolumn{1}{c}{---} & 0.271 & 0.007 & SF      & (1)\\
UGC 07179 & 12:10:58.08 & +63:54:52.7 & 0.175 & \multicolumn{1}{c}{---} & 0.009 & Composite    & (1)\\
CGCG 215-050 NED02 & 12:11:56.48 & +40:39:18.5 & 0.067 & 0.352 & 0.023 & Unknown & \multicolumn{1}{c}{---}\\
NGC 4175 & 12:12:31.05 & +29:10:05.9 & 0.179 & 0.265 & 0.014 & Type 2     & (16)\\
NGC 4194 & 12:14:09.62 & +54:31:36.0 & 0.623 & 1.934 & 0.008 & Type 1     & (1)\\
PG 1211+143 & 12:14:17.77 & +14:03:12.8 & 0.089 & 0.217 & 0.081 & Type 1     & (2),(3)\\
CGCG 013-111 & 12:14:51.29 & -03:29:22.4 & \multicolumn{1}{c}{---} & 0.243 & 0.033 & Type 2     & (1)\\
NGC 4226 & 12:16:26.29 & +47:01:32.4 & 0.124 & \multicolumn{1}{c}{---} & 0.024 & Unknown & \multicolumn{1}{c}{---}\\
IC 0773 & 12:18:08.09 & +06:08:22.4 & \multicolumn{1}{c}{---} & 0.257 & 0.018 & Composite    & (1)\\
NGC 4253 & 12:18:26.54 & +29:48:47.2 & 0.220 & 0.859 & 0.013 & Composite    & (1)\\
SDSS J121849.71+142458.3 & 12:18:49.88 & +14:24:59.4 & 1.428 & \multicolumn{1}{c}{---} & 0.008 & SF      & (2)\\
SDSS J121956.14+052039.5 & 12:19:56.13 & +05:20:39.0 & 0.600 & 0.662 & 0.008 & SF      & (1)\\
VCC 0435 & 12:20:47.22 & +17:00:58.3 & 0.105 & 0.186 & 0.026 & SF      & (1)\\
NGC 4290 & 12:20:47.55 & +58:05:32.9 & 0.135 & 0.281 & 0.010 & SF      & (1)\\
NGC 4332 & 12:22:46.76 & +65:50:37.6 & 0.189 & 0.417 & 0.009 & Composite    & (1)\\
NGC 4334 & 12:23:23.82 & +07:28:23.2 & 0.190 & 0.254 & 0.014 & SF      & (1)\\
SDSS J122512.25+543019.4 & 12:25:11.93 & +54:30:19.1 & 0.158 & \multicolumn{1}{c}{---} & 0.008 & SF      & (1)\\
NGC 4385 & 12:25:42.75 & +00:34:21.8 & 0.167 & 0.635 & 0.007 & SF      & (4)\\
NGC 4388 & 12:25:46.83 & +12:39:43.1 & 0.462 & 1.589 & 0.009 & Type 2     & (1)\\
NGC 4355 & 12:26:54.64 & -00:52:39.3 & 0.266 & 3.286 & 0.007 & Type 2     & (3)\\
NGC 4441 & 12:27:20.20 & +64:48:05.8 & \multicolumn{1}{c}{---} & 0.212 & 0.009 & Composite    & (1)\\
KUG 1225+404 & 12:27:38.00 & +40:09:38.2 & \multicolumn{1}{c}{---} & 0.220 & 0.037 & Composite    & (1)\\
3C 273 & 12:29:06.77 & +02:03:08.0 & 0.276 & 0.454 & 0.158 & Type 1     & (2)\\
NGC 4500 & 12:31:22.22 & +57:57:51.9 & 0.173 & 0.320 & 0.010 & SF      & (3)\\
NGC 4495 & 12:31:22.84 & +29:08:11.1 & 0.205 & \multicolumn{1}{c}{---} & 0.015 & Composite    & (1)\\
N4532 & 12:34:19.24 & +06:28:08.9 & \multicolumn{1}{c}{---} & 0.263 & 0.007 & SF      & (2)\\
NGC 4535 & 12:34:20.39 & +08:11:52.7 & 0.088 & 0.276 & 0.007 & SF      & (11)\\
NGC 4536 & 12:34:27.08 & +02:11:18.0 & 1.018 & \multicolumn{1}{c}{---} & 0.006 & SF      & (11)\\
NGC 4568 & 12:36:34.25 & +11:14:19.4 & 0.993 & 0.914 & 0.007 & SF      & (1)\\
IC 3581 & 12:36:37.98 & +24:25:43.0 & \multicolumn{1}{c}{---} & 0.260 & 0.023 & Composite    & (1)\\
KUG 1238+278A & 12:40:46.40 & +27:33:54.2 & \multicolumn{1}{c}{---} & 0.240 & 0.056 & LINER   & (1)\\
WAS 61 & 12:42:10.68 & +33:17:02.8 & \multicolumn{1}{c}{---} & 0.222 & 0.044 & Type 1     & (1)\\
MCG +07-26-051 & 12:46:56.82 & +42:16:00.2 & 0.161 & 0.219 & 0.033 & Unknown & \multicolumn{1}{c}{---}\\
NGC 4704 & 12:48:46.46 & +41:55:16.7 & \multicolumn{1}{c}{---} & 0.227 & 0.027 & Type 2     & (3)\\
2MASX J12494552+4328570 & 12:49:45.51 & +43:28:56.2 & \multicolumn{1}{c}{---} & 0.199 & 0.062 & Type 2     & (1)\\
2MASX J12501385+0734443 & 12:50:13.81 & +07:34:44.7 & 0.098 & 0.239 & 0.038 & SF      & (1)\\
SHOC 391 & 12:53:05.97 & -03:12:58.7 & \multicolumn{1}{c}{---} & 0.298 & 0.023 & Composite    & (1)\\
NGC 4793 & 12:54:40.94 & +28:56:20.7 & 0.809 & \multicolumn{1}{c}{---} & 0.008 & SF      & (1)\\
IC 0836 & 12:55:54.22 & +63:36:41.9 & 0.095 & 0.151 & 0.009 & Unknown & \multicolumn{1}{c}{---}\\
UGC 08058 & 12:56:14.27 & +56:52:25.4 & 1.051 & 4.396 & 0.042 & Type 1     & (2)\\
2MASX J13000533+1632151 & 13:00:05.31 & +16:32:14.2 & 0.088 & 0.175 & 0.080 & Type 2     & (1)\\
NGC 4922 NED02 & 13:01:25.28 & +29:18:50.4 & 0.149 & 0.740 & 0.023 & Type 2     & (1)\\
VV 283a & 13:01:50.38 & +04:19:58.9 & 0.107 & 0.322 & 0.037 & Type 2     & (1)\\
NGC 4963 & 13:05:52.04 & +41:43:18.2 & 0.101 & \multicolumn{1}{c}{---} & 0.024 & Unknown & \multicolumn{1}{c}{---}\\
UGC 08237 & 13:08:54.20 & +62:18:23.1 & 0.122 & \multicolumn{1}{c}{---} & 0.010 & SF      & (10)\\
NGC 5020 & 13:12:39.88 & +12:35:59.8 & 0.103 & \multicolumn{1}{c}{---} & 0.011 & SF      & (10)\\
IC 0860 & 13:15:03.49 & +24:37:08.1 & \multicolumn{1}{c}{---} & 0.285 & 0.013 & Unknown & \multicolumn{1}{c}{---}\\
UGC 08327 NED02 & 13:15:17.27 & +44:24:26.6 & 0.092 & 0.264 & 0.035 & LINER   & (1)\\
UGC 08335 NED02 & 13:15:34.91 & +62:07:29.0 & 0.168 & 0.853 & 0.031 & Composite    & (1)\\
2MASX J13163979+4452351 & 13:16:39.69 & +44:52:35.3 & 0.135 & 0.360 & 0.091 & Type 2     & (1)\\
NGC 5060 & 13:17:16.22 & +06:02:14.4 & \multicolumn{1}{c}{---} & 0.189 & 0.021 & Unknown & \multicolumn{1}{c}{---}\\
IC 0883 & 13:20:35.31 & +34:08:22.1 & 0.213 & 0.554 & 0.023 & Composite    & (1)\\
NGC 5104 & 13:21:23.11 & +00:20:32.7 & 0.204 & 0.307 & 0.019 & Composite    & (12)\\
$[HB89]$ 1321+058 & 13:24:19.88 & +05:37:04.4 & 0.216 & 0.328 & 0.203 & Type 1     & (1)\\
NGC 5149 & 13:26:09.16 & +35:56:04.5 & 0.115 & 0.291 & 0.019 & Composite    & (1)\\
2MASX J13315286+0200596 & 13:31:52.83 & +02:00:59.7 & \multicolumn{1}{c}{---} & 0.170 & 0.086 & Type 2     & (1)\\
NGC 5218 & 13:32:10.30 & +62:46:04.1 & 0.188 & 0.484 & 0.010 & Composite    & (17)\\
SDSS J133223.99+110620.3 & 13:32:24.05 & +11:06:19.8 & 0.074 & 0.282 & 0.031 & SF      & (1)\\
UGC 08561 & 13:34:57.36 & +34:02:39.0 & 0.174 & \multicolumn{1}{c}{---} & 0.024 & SF      & (1)\\
$[HB89]$ 1334+246 & 13:37:18.71 & +24:23:02.8 & 0.449 & 0.612 & 0.108 & Type 1     & (2),(3)\\
SDSS J133817.27+481632.2 & 13:38:17.26 & +48:16:33.3 & 0.165 & 0.556 & 0.028 & Type 2     & (1)\\
NGC 5257 & 13:39:52.94 & +00:50:23.6 & 0.295 & \multicolumn{1}{c}{---} & 0.023 & SF      & (2)\\
NGC 5263 & 13:39:55.56 & +28:24:01.5 & 0.126 & \multicolumn{1}{c}{---} & 0.016 & SF      & (18)\\
NGC 5258 & 13:39:57.49 & +00:49:48.0 & 0.299 & 0.311 & 0.023 & SF      & (4)\\
CGCG 045-068 & 13:40:27.15 & +04:46:24.8 & 0.075 & 0.250 & 0.023 & Composite    & (1)\\
IC 0910 & 13:41:07.82 & +23:16:55.3 & \multicolumn{1}{c}{---} & 0.265 & 0.027 & LINER   & (2)\\
MRK 0268 & 13:41:11.09 & +30:22:41.0 & \multicolumn{1}{c}{---} & 0.211 & 0.040 & Type 1     & (1)\\
MRK 0273:$[XXM2002]$ Radio N & 13:44:42.09 & +55:53:13.7 & 0.135 & 0.871 & 0.037 & Type 2     & (1)\\
MRK 0796 & 13:46:49.45 & +14:24:00.9 & 0.116 & 0.429 & 0.021 & Composite    & (1)\\
MRK 1361 & 13:47:04.36 & +11:06:22.5 & 0.141 & 0.398 & 0.023 & Type 2     & (1)\\
2MASX J13470695+3456238 & 13:47:06.98 & +34:56:24.9 & \multicolumn{1}{c}{---} & 0.239 & 0.054 & Composite    & (1)\\
4C +12.50 & 13:47:33.33 & +12:17:24.8 & \multicolumn{1}{c}{---} & 0.311 & 0.121 & Type 1     & (1)\\
UGC 08728 & 13:48:12.71 & +07:23:42.5 & 0.115 & 0.179 & 0.023 & Type 2     & (1)\\
FIRST J134913.9+351525 & 13:49:14.18 & +35:15:23.8 & 0.251 & \multicolumn{1}{c}{---} & 0.017 & SF      & (18)\\
NGC 5331 NED01 & 13:52:16.18 & +02:06:05.7 & 0.192 & \multicolumn{1}{c}{---} & 0.033 & SF      & (19)\\
PG 1351+640 & 13:53:15.80 & +63:45:45.6 & 0.108 & 0.307 & 0.088 & Type 1     & (2),(3)\\
NGC 5347 & 13:53:17.85 & +33:29:26.9 & 0.138 & 0.601 & 0.008 & Type 2     & (1)\\
UGC 08827 & 13:54:31.12 & +15:02:39.3 & 0.160 & 0.362 & 0.018 & Composite    & (1)\\
MRK 0463E & 13:56:02.86 & +18:22:18.4 & 0.376 & 1.103 & 0.051 & Type 2     & (1)\\
NGC 5383 & 13:57:04.94 & +41:50:47.2 & 0.262 & 0.346 & 0.007 & SF      & (1)\\
NGC 5374 & 13:57:29.66 & +06:05:48.0 & 0.123 & \multicolumn{1}{c}{---} & 0.015 & SF      & (1)\\
SDSS J135817.50+010843.4 & 13:58:17.47 & +01:08:44.6 & \multicolumn{1}{c}{---} & 0.285 & 0.035 & SF      & (1)\\
NGC 5394 & 13:58:33.65 & +37:27:11.8 & 0.237 & 0.413 & 0.012 & Composite    & (1)\\
MCG +06-31-036 & 13:58:41.85 & +35:05:15.4 & 0.105 & 0.175 & 0.035 & SF      & (1)\\
NGC 5430 & 14:00:45.78 & +59:19:42.5 & 0.375 & 0.813 & 0.010 & SF      & (1)\\
NGC 5433 & 14:02:36.08 & +32:30:37.1 & 0.230 & 0.360 & 0.015 & SF      & (1)\\
$[HB89]$ 1402+436 & 14:04:38.68 & +43:27:07.3 & \multicolumn{1}{c}{---} & 0.254 & 0.323 & Type 1     & (1)\\
VIII Zw 357 & 14:06:09.50 & +02:19:52.8 & 0.119 & 0.319 & 0.025 & SF      & (1)\\
NGC 5480 & 14:06:21.42 & +50:43:28.3 & 0.219 & \multicolumn{1}{c}{---} & 0.006 & SF      & (1)\\
MRK 0668 & 14:07:00.43 & +28:27:15.0 & 0.080 & 0.305 & 0.077 & Type 1     & (1)\\
CGCG 074-129 & 14:10:41.41 & +13:33:28.9 & \multicolumn{1}{c}{---} & 0.471 & 0.016 & Type 2     & (1)\\
CGCG 018-077 & 14:12:15.63 & -00:38:00.4 & 0.091 & \multicolumn{1}{c}{---} & 0.026 & Unknown & \multicolumn{1}{c}{---}\\
NGC 5506 & 14:13:14.86 & -03:12:26.8 & 0.823 & 2.240 & 0.006 & Type 2     & (1)\\
NGC 5541 & 14:16:31.77 & +39:35:25.3 & 0.234 & \multicolumn{1}{c}{---} & 0.026 & SF      & (18)\\
IC 4395 & 14:17:20.96 & +26:51:26.1 & 0.119 & 0.178 & 0.037 & Composite    & (1)\\
NGC 5548 & 14:17:59.58 & +25:08:12.6 & 0.157 & 0.409 & 0.017 & Type 1     & (1)\\
UGC 09165 & 14:18:47.65 & +24:56:21.4 & 0.232 & \multicolumn{1}{c}{---} & 0.018 & SF      & (18)\\
MRK 1490 & 14:19:43.20 & +49:14:10.8 & 0.088 & 0.395 & 0.026 & Composite    & (1)\\
IC 4408 & 14:21:12.95 & +29:59:36.1 & 0.113 & \multicolumn{1}{c}{---} & 0.030 & Composite    & (1)\\
SBS 1419+480 & 14:21:29.78 & +47:47:25.3 & 0.046 & 0.189 & 0.073 & Type 1     & (1)\\
NGC 5600 & 14:23:49.39 & +14:38:19.7 & 0.348 & \multicolumn{1}{c}{---} & 0.008 & Unknown & \multicolumn{1}{c}{---}\\
NGC 5610 & 14:24:22.95 & +24:36:51.8 & 0.167 & 0.371 & 0.017 & Composite    & (1)\\
NGC 5633 & 14:27:28.53 & +46:08:47.1 & 0.308 & \multicolumn{1}{c}{---} & 0.008 & SF      & (1)\\
NGC 5653 & 14:30:10.29 & +31:12:56.3 & 0.605 & 0.799 & 0.012 & SF      & (21)\\
NGC 5678:$[HK83]$ 42 & 14:32:05.06 & +57:55:16.8 & 0.419 & \multicolumn{1}{c}{---} & 0.006 & SF      & (3)\\
SDSS J143225.99+080448.4 & 14:32:25.91 & +08:04:46.9 & 0.332 & 0.513 & 0.007 & SF      & (1)\\
SDSS J143247.74+492751.5 & 14:32:47.59 & +49:27:48.2 & \multicolumn{1}{c}{---} & 0.384 & 0.006 & SF      & (1)\\
KUG 1433+353 & 14:35:18.23 & +35:07:08.2 & 0.134 & \multicolumn{1}{c}{---} & 0.029 & Composite    & (1)\\
UGC 09412 & 14:36:22.10 & +58:47:39.9 & 0.188 & 0.669 & 0.031 & Type 1     & (2),(3)\\
UGC 09425 NED02 & 14:37:51.08 & +30:28:48.7 & \multicolumn{1}{c}{---} & 0.179 & 0.035 & SF      & (10)\\
SDSS J144011.31-001719.5 & 14:40:11.35 & -00:17:22.6 & 1.010 & 1.513 & 0.006 & SF      & (4)\\
MRK 0477 & 14:40:38.11 & +53:30:15.8 & \multicolumn{1}{c}{---} & 0.352 & 0.038 & Type 2     & (1)\\
2MASX J14410437+5320088 & 14:41:04.38 & +53:20:09.5 & \multicolumn{1}{c}{---} & 0.205 & 0.105 & Type 2     & (1)\\
2MASX J14492067+4221013 & 14:49:20.75 & +42:21:00.6 & 0.093 & 0.162 & 0.179 & Type 1     & (1)\\
NGC 5765B & 14:50:51.45 & +05:06:52.2 & 0.147 & 0.408 & 0.028 & Type 2     & (1)\\
UGC 09561 NOTES01 & 14:51:29.35 & +09:20:05.2 & 0.114 & 0.164 & 0.029 & Composite    & (1)\\
N5795 & 14:56:19.32 & +49:23:55.2 & 0.153 & \multicolumn{1}{c}{---} & 0.008 & Unknown & \multicolumn{1}{c}{---}\\
UGC 09618 NED02 & 14:57:00.69 & +24:37:04.5 & 0.240 & 0.289 & 0.034 & SF      & (22)\\
NGC 5804 & 14:57:06.90 & +49:40:09.5 & 0.115 & 0.108 & 0.013 & Type 1     & (1)\\
NGC 5792 & 14:58:22.67 & -01:05:28.1 & 0.330 & 0.472 & 0.006 & SF      & (23)\\
UGC 09639 & 14:58:36.05 & +44:53:00.6 & 0.158 & 0.297 & 0.036 & Type 2     & (1)\\
CGCG 105-099 & 15:02:53.23 & +16:55:09.0 & \multicolumn{1}{c}{---} & 0.263 & 0.021 & Composite    & (1)\\
MRK 0841 & 15:04:01.21 & +10:26:15.6 & 0.126 & 0.372 & 0.036 & Type 1     & (2),(3)\\
2MASX J15051788+0809127 & 15:05:17.93 & +08:09:13.4 & \multicolumn{1}{c}{---} & 0.188 & 0.039 & Composite    & (1)\\
NGC 5860 NED01 & 15:06:33.57 & +42:38:25.6 & 0.093 & \multicolumn{1}{c}{---} & 0.018 & SF      & (1)\\
VV 059a & 15:08:05.69 & +34:23:21.3 & 0.096 & \multicolumn{1}{c}{---} & 0.046 & Unknown & \multicolumn{1}{c}{---}\\
CGCG 077-021 & 15:09:08.72 & +09:02:20.1 & \multicolumn{1}{c}{---} & 0.228 & 0.044 & SF      & (3)\\
NGC 5875 & 15:09:12.93 & +52:31:41.0 & 0.164 & \multicolumn{1}{c}{---} & 0.012 & Type 2     & (1)\\
2MASX J15114125+0518089 & 15:11:41.28 & +05:18:08.4 & \multicolumn{1}{c}{---} & 0.249 & 0.085 & Type 1     & (1)\\
CGCG 049-057 & 15:13:13.21 & +07:13:32.7 & \multicolumn{1}{c}{---} & 0.179 & 0.013 & Composite    & (24)\\
NGC 5900 & 15:15:05.29 & +42:12:35.0 & 0.343 & 0.385 & 0.008 & Unknown & \multicolumn{1}{c}{---}\\
NGC 5905 & 15:15:23.45 & +55:31:03.3 & 0.133 & 0.137 & 0.011 & SF      & (3)\\
SDSS J151548.37+200121.5 & 15:15:48.36 & +20:01:22.2 & 0.066 & 0.181 & 0.036 & SF      & (1)\\
2MASSi J1516532+190048 & 15:16:53.25 & +19:00:47.8 & 0.102 & \multicolumn{1}{c}{---} & 0.190 & Type 1     & (2),(3)\\
VV 705 NED01 & 15:18:06.13 & +42:44:44.5 & 0.156 & 0.645 & 0.040 & Composite    & (1)\\
SBS 1517+522 & 15:19:07.47 & +52:06:06.9 & \multicolumn{1}{c}{---} & 0.182 & 0.138 & Type 1     & (1)\\
SDSS J151947.03+394534.8 & 15:19:47.07 & +39:45:36.4 & \multicolumn{1}{c}{---} & 0.266 & 0.047 & SF      & (1)\\
SDSS J152238.10+333135.8 & 15:22:38.05 & +33:31:35.2 & \multicolumn{1}{c}{---} & 0.194 & 0.125 & Composite    & (1)\\
2MASX J15234787+2855032 & 15:23:47.83 & +28:55:02.4 & \multicolumn{1}{c}{---} & 0.227 & 0.087 & Type 1     & (1)\\
NGC 5930 & 15:26:07.88 & +41:40:33.7 & 0.227 & 0.782 & 0.009 & Composite    & (1)\\
CGCG 021-096 & 15:26:37.63 & +00:35:32.6 & 0.065 & 0.220 & 0.051 & Type 2     & (1)\\
IRAS 15250+3609 & 15:26:59.46 & +35:58:37.2 & 0.098 & 0.452 & 0.055 & Composite    & (1)\\
NGC 5936 & 15:30:00.85 & +12:59:20.2 & 0.386 & 0.663 & 0.013 & Composite    & (1)\\
NGC 5937 & 15:30:46.17 & -02:49:44.9 & 0.597 & 0.843 & 0.009 & SF      & (16)\\
2MASX J15322966+3007490 & 15:32:29.78 & +30:07:48.4 & \multicolumn{1}{c}{---} & 0.222 & 0.065 & Type 2     & (1)\\
NGC 5953 & 15:34:32.36 & +15:11:37.5 & 0.408 & 0.580 & 0.007 & Type 2     & (3)\\
NGC 5954 & 15:34:35.07 & +15:12:00.8 & 0.197 & 0.299 & 0.006 & SF      & (1)\\
ARP 220:$[LDT2006]$ E13 & 15:34:57.24 & +23:30:11.6 & 0.249 & 2.261 & 0.018 & Type 1     & (1)\\
NGC 5962 & 15:36:31.80 & +16:36:27.0 & 0.366 & \multicolumn{1}{c}{---} & 0.007 & Unknown & \multicolumn{1}{c}{---}\\
SDSS J153927.51+245651.4 & 15:39:27.49 & +24:56:51.0 & \multicolumn{1}{c}{---} & 0.235 & 0.023 & SF      & (3)\\
NGC 5975 & 15:39:57.95 & +21:28:13.2 & 0.166 & 0.188 & 0.015 & Type 2     & (1)\\
NGC 5980 & 15:41:30.49 & +15:47:18.9 & 0.274 & \multicolumn{1}{c}{---} & 0.014 & Unknown & \multicolumn{1}{c}{---}\\
CGCG 166-047 & 15:43:57.35 & +28:31:26.5 & \multicolumn{1}{c}{---} & 0.180 & 0.032 & Type 2     & (1)\\
NGC 5990 & 15:46:16.41 & +02:24:55.3 & 0.450 & 0.802 & 0.012 & SF      & (1)\\
UGC 10029 & 15:46:54.32 & +05:53:27.7 & 0.097 & 0.196 & 0.042 & Unknown & \multicolumn{1}{c}{---}\\
UGC 10070 & 15:51:12.88 & +47:15:17.6 & 0.255 & 0.327 & 0.019 & Unknown & \multicolumn{1}{c}{---}\\
MRK 0492 & 15:58:43.72 & +26:49:04.2 & 0.092 & 0.370 & 0.014 & Composite    & (1)\\
NGC 6052 NED01 & 16:05:12.93 & +20:32:35.4 & 0.200 & 0.397 & 0.016 & SF      & (1)\\
GIN 510 & 16:07:38.49 & +18:28:51.1 & \multicolumn{1}{c}{---} & 0.176 & 0.037 & Composite    & (1)\\
IC 1198 & 16:08:36.37 & +12:19:49.8 & \multicolumn{1}{c}{---} & 0.246 & 0.034 & Type 1     & (1)\\
2MASX J16094822+0434525 & 16:09:48.24 & +04:34:52.8 & \multicolumn{1}{c}{---} & 0.248 & 0.064 & Type 2     & (1)\\
NGC 6090 NED02 & 16:11:40.77 & +52:27:27.0 & 0.196 & 0.499 & 0.029 & SF      & (1)\\
UGC 10322 & 16:18:07.93 & +22:13:32.3 & 0.183 & \multicolumn{1}{c}{---} & 0.015 & SF      & (1)\\
NGC 6120 & 16:19:48.05 & +37:46:28.5 & 0.164 & 0.270 & 0.031 & SF      & (25)\\
SBS 1620+504 & 16:22:17.74 & +50:22:20.1 & 0.116 & 0.164 & 0.057 & Composite    & (1)\\
UGC 10384 & 16:26:46.63 & +11:34:49.9 & 0.122 & \multicolumn{1}{c}{---} & 0.017 & SF      & (1)\\
SN 1951I & 16:32:20.81 & +19:49:32.8 & \multicolumn{1}{c}{---} & 0.556 & 0.008 & SF      & (11)\\
SDSS J163425.28+213228.5 & 16:34:25.37 & +21:32:27.7 & 0.162 & 0.201 & 0.010 & SF      & (1)\\
FBQS J164018.1+384220 & 16:40:18.18 & +38:42:20.4 & \multicolumn{1}{c}{---} & 0.233 & 0.020 & PN      & (2),(3)\\
3C 345 & 16:42:58.88 & +39:48:37.2 & 0.102 & 0.353 & 0.593 & Type 1     & (2),(3)\\
NGC 6247 & 16:48:20.30 & +62:58:36.5 & 0.233 & 0.289 & 0.015 & SF      & (1)\\
2MASX J16512188+2155264 & 16:51:21.89 & +21:55:26.5 & 0.119 & \multicolumn{1}{c}{---} & 0.055 & Type 2     & (1)\\
NGC 6306 & 17:07:36.90 & +60:43:43.1 & 0.095 & 0.232 & 0.010 & SF      & (10)\\
NGC 6361 & 17:18:41.14 & +60:36:28.6 & 0.266 & \multicolumn{1}{c}{---} & 0.013 & Unknown & \multicolumn{1}{c}{---}\\
2MASX J17380143+5613257 & 17:38:01.37 & +56:13:26.2 & 0.070 & 0.241 & 0.065 & Type 2     & (1)\\
MCG -01-53-004 & 20:45:14.66 & -05:29:03.1 & \multicolumn{1}{c}{---} & 0.337 & 0.027 & Composite    & (1)\\
NGC 6967 & 20:47:34.10 & +00:24:40.0 & \multicolumn{1}{c}{---} & 0.285 & 0.013 & Type 2     & (2),(3)\\
UGC 11763 & 21:32:27.84 & +10:08:18.8 & 0.134 & 0.236 & 0.063 & Type 1     & (2),(3)\\
2MASX J21522605-0810248 & 21:52:26.05 & -08:10:25.7 & \multicolumn{1}{c}{---} & 0.324 & 0.035 & Type 2     & (1)\\
UGC 11865 & 21:58:36.01 & +12:02:20.6 & 0.142 & 0.197 & 0.031 & SF      & (26)\\
NGC 7189 & 22:03:16.00 & +00:34:15.5 & \multicolumn{1}{c}{---} & 0.197 & 0.030 & Type 2     & (1)\\
CGCG 429-014 & 22:37:07.77 & +14:13:52.8 & 0.116 & \multicolumn{1}{c}{---} & 0.038 & Composite    & (1)\\
NGC 7364 & 22:44:24.39 & -00:09:41.1 & 0.106 & \multicolumn{1}{c}{---} & 0.016 & Unknown & \multicolumn{1}{c}{---}\\
2MASX J22533299-0024428 & 22:53:32.93 & -00:24:42.0 & \multicolumn{1}{c}{---} & 0.311 & 0.058 & Composite    & (1)\\
IC 1464B & 23:03:11.18 & -08:59:23.6 & 0.118 & 0.231 & 0.024 & SF      & (26),(27)\\
MRK 0926 & 23:04:43.62 & -08:41:09.1 & 0.060 & 0.214 & 0.047 & Type 1     & (1)\\
UGC 12348 & 23:05:18.79 & +00:11:23.1 & 0.114 & 0.172 & 0.025 & Type 2     & (3)\\
NGC 7580 & 23:17:36.44 & +14:00:04.9 & 0.137 & \multicolumn{1}{c}{---} & 0.015 & SF      & (1)\\
NGC 7603 & 23:18:56.68 & +00:14:37.5 & 0.295 & 0.321 & 0.030 & Type 1     & (2),(3)\\
UGC 12633 & 23:30:13.70 & +15:45:40.3 & 0.127 & 0.190 & 0.014 & Composite    & (1)\\
NGC 7738 & 23:44:02.07 & +00:31:00.5 & 0.128 & 0.264 & 0.023 & SF      & (4)\\
2MASX J23470918+1535487 & 23:47:09.19 & +15:35:48.5 & \multicolumn{1}{c}{---} & 0.571 & 0.026 & Type 2     & (2)\\
\end{longtable}
\end{small}

\subsection{Classification of Spectroscopic Galaxy Type}
\label{Classification}
\begin{figure}
  \begin{center}
    \FigureFile(80mm,50mm){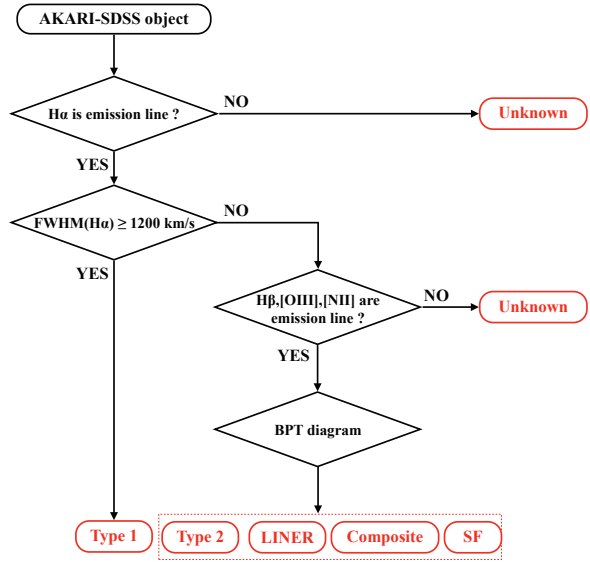}
  \end{center}
\caption{Outline of the type classification process.}
\label{SDSS_type_classification}
\end{figure}

We classified the 243 and 255 galaxies into the following five types by using the line width of H$\alpha$ or the emission line ratios of [OIII]/H$\beta$ and [NII]/H$\alpha$, as shown in figure \ref{SDSS_type_classification}: Type 1 AGNs (Type1); Type 2 AGNs (Type 2); low-ionization narrow emission line galaxies (LINER); galaxies that are likely to contain both star formation and AGN activity (composite types of galaxies, hereinafter Composite); and star-forming galaxies (SF).

\begin{figure}
  \begin{center}
    \FigureFile(80mm,50mm){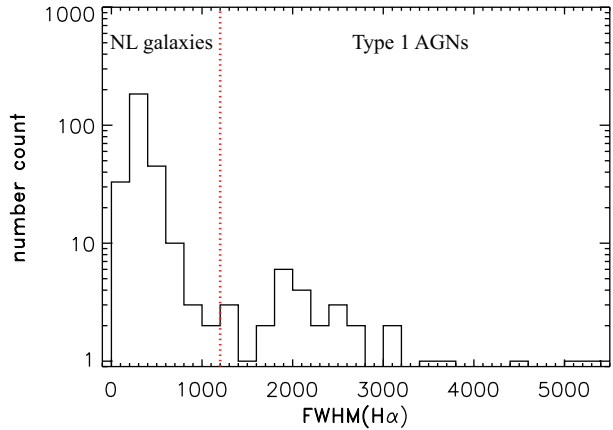}
  \end{center}
\caption{Distribution of the full-width at half-maximum (FWHM) of $H\alpha$ for AKARI-SDSS galaxies. We selected objects with an FWHM(H$\alpha$) greater than 1200 km s$^{-1}$ as Type 1 AGNs. Narrow emission line galaxies (NL galaxies) were defined as having an FWHM(H$\alpha$) smaller than 1200 km s$^{-1}$ in this study.}
\label{FWHM}
\end{figure}

\begin{figure}
  \begin{center}
    \FigureFile(80mm,50mm){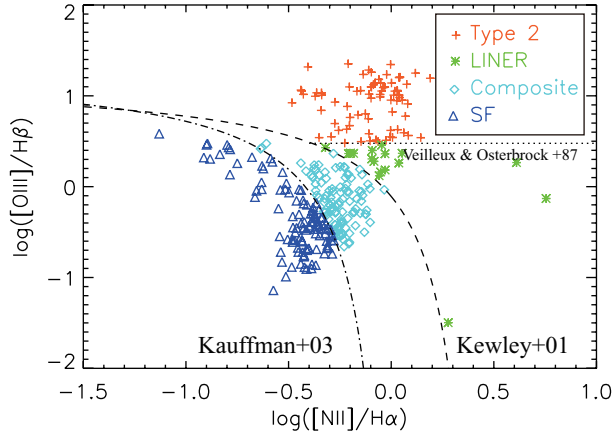}
  \end{center}
\caption{BPT diagram plotting the emission-line flux ratio [NII]/H$\alpha$ versus [OIII]/H$\beta$ for all the NL galaxies in which the line flux information is available. The dashed-dotted line is the criterion given by \citet{Kauffmann}, the dashed line is the criterion given by \citet{Kewley+01}, and the dotted line is the traditional scheme (see for example, \cite{Veilleux+87}).}
\label{BPT}
\end{figure}

First, we separated the Type 1 objects using the full-width at half-maximum (FWHM) of the H$\alpha$ emission line. The FWHM was estimated by the following relation:
\begin{equation}
FWHM(H\alpha) = \frac{2.35 \sigma c}{\lambda_0 (1+z)},
\end{equation}
where $\sigma$ is the variance of the Gaussian curve that fits the H$\alpha$ emission line, $c$ is the speed of light, and $\lambda_0$ is the rest-frame wavelength of the H$\alpha$ line (6563 \AA).
 The values of $\sigma$ and $z$ were obtained from the ``sigma'' and ``z'' columns, respectively, in the SDSS DR7 {\tt SpecLine} table.
 For ZCAT galaxies, we classified Type 1 according to the previous literature, because the ZCAT does not contain spectroscopic line information. 
 Figure \ref{FWHM} gives the distribution of the FWHM values of the H$\alpha$ emission line. 
 The distribution is bimodal with a minimum at 1200 km s$^{-1}$, as reported by \citet{Haoa}.
 They established the criterion of FWHM(H$\alpha$) $\geq$ 1200 km s$^{-1}$ for identifying broad-line AGNs from spectroscopic SDSS data. 
 Similarly, we extracted objects with an FWHM(H$\alpha$) greater than 1200 km s$^{-1}$ as Type 1.
 Next, objects that have H$\alpha$ emission with an FWHM $\leq$ 1200 km $^{-1}$ were classified as Type 2, LINER, Composite, or SF by using the optical flux line ratios of [OIII]$\lambda$5007/H$\beta$ versus [NII]$\lambda$6583/H$\alpha$ (BPT diagram suggested by \cite{Baldwin}), as shown in figure \ref{BPT}.
 These line fluxes were estimated by $\sqrt{2 \pi} \sigma h$, where $h$ is the height of the Gaussian curve that fits each emission line. 
 The $h$ values were obtained from the ``height'' column in the SDSS DR7 {\tt SpecLine} table. 
 In this study, we did not correct the effect of stellar absorption. 
 The ZCAT galaxies were also classified as Type 2, LINER, SF, or Composite according to previous literature.
However, some galaxies were not classified, because (i) H$\alpha$, H$\beta$, [OIII] or [NII] were not detected (neither in absorption nor emission) in the SDSS, and (ii) no spectroscopic type was available in the literature.
 These galaxies are classified as unknown types of galaxies (hereinafter, ``Unknown'').
 
 The final classification of each of the 243 and 255 galaxies are summarized in table \ref{type_classification}. 
 As shown in table \ref{type_classification}, the 9 $\micron$ LF sample has 25 Type 1s, 39 Type 2s, 6 LINERs, 50 Composites, 95 SFs, and 28 Unknowns.
 The 18 $\micron$ LF sample has 41 Type 1s, 64 Type 2s, 11 LINERs, 57 Composites, 71 SFs, and 11 Unknowns.
 These results strongly confirm the finding of \citet{Spinoglio} that MIR selection strongly favors finding galaxies with active nuclei because of their strong nuclear emissions from heated dust grains.
 In their pioneering study using IRAS, they found that 15\% of 12 $\micron$-selected galaxies are AGNs.
 Our selection using AKARI is even more efficient, containing $\sim$ 26\% and 41\% AGNs (Type 1 + Type 2 AGNs) in the 9 and 18 $\micron$ LF samples, respectively. 
 This is due to AKARI's improved ability to isolate nuclear point-source emission in the MIR.
 Spinoglio \& Malkan's 12 $\micron$ sample, on the other hand, was based on {\it total} fluxes measured by IRAS, which typically include emissions over a few square arcminutes. 
 This much coarser spatial resolution substantially dilutes the infrared signal of the active nucleus. 
 A secondary disadvantage of the IRAS 12 $\micron$ bandpass was that it included a strong potential for contributions from polycyclic aromatic hydrocarbon (PAH) emissions, which are unrelated to the presence of an active nucleus.

\begin{figure*}
   \begin{center}
      \FigureFile(160mm,50mm){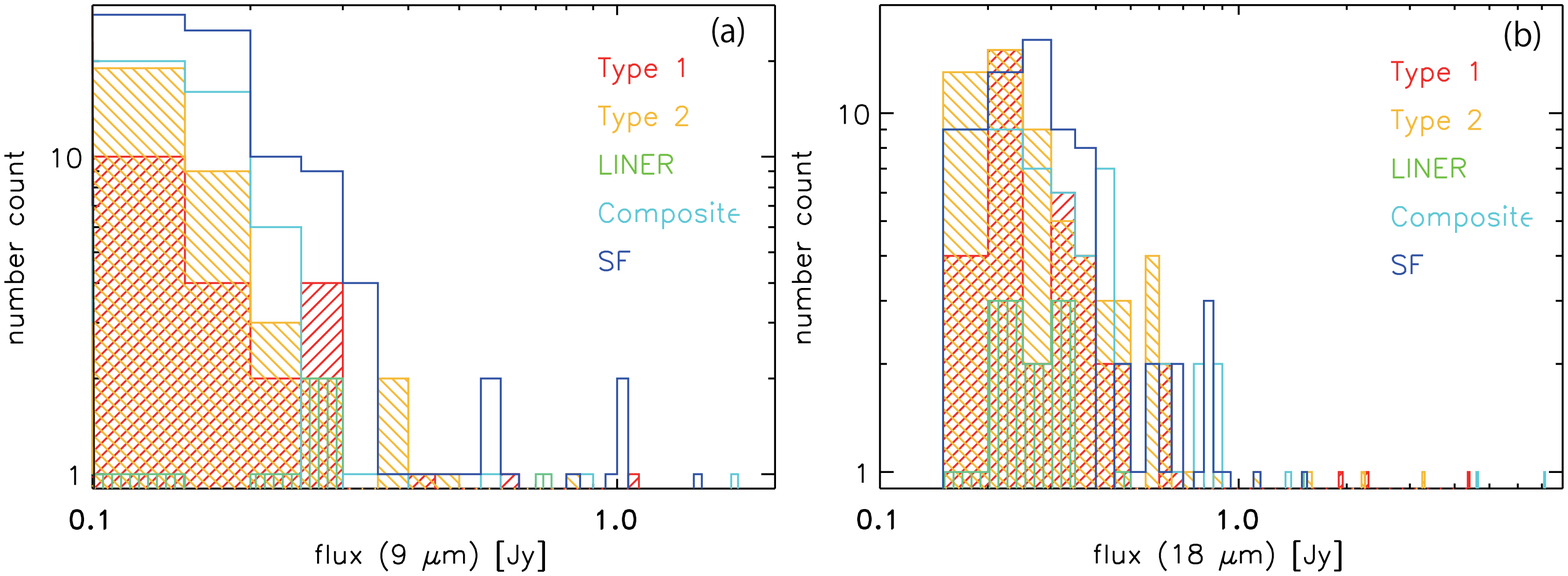}
   \end{center}
\caption{Flux distribution for each type of galaxy at (a) 9 $\micron$ and (b) 18 $\micron$.}
\label{flux}
\end{figure*}

\begin{figure*}
   \begin{center}
      \FigureFile(160mm,50mm){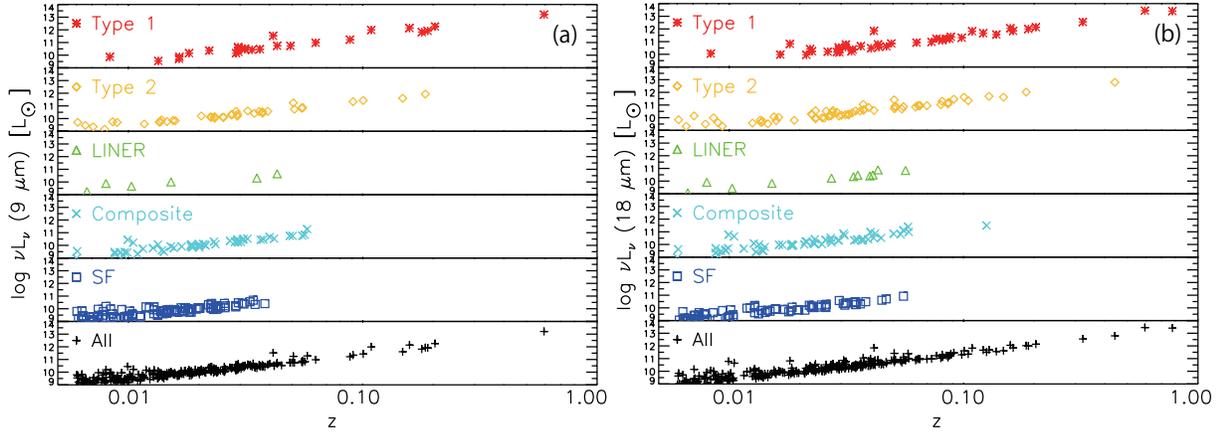}
   \end{center}
\caption{(a) 9 $\micron$ and (b) 18 $\micron$ luminosities as a function of redshift for each type of galaxy.}
\label{L}
\end{figure*}

Table \ref{type_classification} also indicates that the 18 $\micron$ band is especially powerful for finding AGNs. 
 The detection rate of AGNs in the 18 $\micron$ band (41\%) is higher than that in the 9 $\micron$ band ($\sim$ 26\%), since the 9 $\micron$ sample is also affected by PAH contributions in the same way as the 12 $\micron$ sample as mentioned above. 
 Figure \ref{flux} presents the flux distributions at 9 $\micron$ and 18 $\micron$.
 The distributions of 9 and 18 $\micron$ luminosities as a function of redshift are illustrated in figure \ref{L}. 
 For the flux distribution, there is no specific difference between each type of galaxy shown in figure \ref{flux}. 
 In contrast, for the redshift distribution, the Type 2s and especially the Type 1s have higher redshifts than the others.
 
 \begin{table}
  \caption{Summary of the classification types of 243 galaxies for the 9 $\micron$ LF and 255 galaxies for the 18 $\micron$ LF.}
  \label{type_classification}
  \begin{center}
    \begin{tabular}{lrr}
      \hline
      type & 9 $\micron$ (percentage) & 18 $\micron$ (percentage) \\
      \hline
		Type 1		&	25 (10.3\%)	&	41 (16.0\%)\\
		Type 2		&	39 (15.6\%) &	64 (25.0\%)\\
		LINER		&	 6 \,\,\,(2.5\%) 	&	11 \,\,(4.3\%)\\
		Composite	&	50 (20.6\%)	&	57 (22.3\%)\\
		SF			&	95 (39.5\%)	&	71 (28.1\%)\\	
		Unknown		&	28 (11.5\%)	&	11 \,\,(4.3\%) \\	
	  \hline
		All			&	243 (100\%)	&	255 (100\%)\\		
      \hline
    \end{tabular}
  \end{center}
\end{table}

\subsection{Completeness Correction}
\label{Completeness}

\subsubsection{$V/V_{\mathrm{max}}$test}
\label{V/Vmax_test}
Before constructing LFs, we performed the standard $V/V_{\mathrm{max}}$ test (\cite{Schmidt}) to examine whether the spatial distribution of the sources in a sample is uniform.
 Here $V$ is the volume enclosed at the redshift of an object, and $V_{\mathrm{max}}$ is the volume that would be enclosed at the maximum redshift at which the object could be detected. 
 If the sample is complete, the mean value of $V/V_{\mathrm{max}}$ should be 0.5.

In the context of the cosmology we adopted, $V$ and $V_{\mathrm{max}}$ are described as
\begin{eqnarray}
V(z) & = & \frac{c}{H_0} \int_{\Omega} \int_{z_{\mathrm{min}}}^{z} C(z^{\prime}) \frac{(1+z^{\prime})^2 D_A^2}{\sqrt{\Omega_M (1+z^{\prime})^3 + \Omega_{\Lambda}}} dz^{\prime} d\Omega, \\
V_{\mathrm{max}}(z) & = & \frac{c}{H_0} \int_{\Omega} \int_{z_{\mathrm{min}}}^{z_{\mathrm{max}}} C(z^{\prime}) \frac{(1+z^{\prime})^2 D_A^2}{\sqrt{\Omega_M (1+z^{\prime})^3 + \Omega_{\Lambda}}} dz^{\prime} d\Omega, 
\end{eqnarray}
where $D_A$ is the angular distance for a given redshift in our adopted cosmology, $\Omega$ is the solid angle of the SDSS spectroscopic region (8032 deg$^2$), and $z_{\mathrm{min}}$ = 0.006 (see section \ref{sample_selection}) is the lower limit of the redshift bin considered.
 $z_{\mathrm{max}}$ is the maximum redshift at which the object could be seen, given the flux limit of the sample. 
 
 However, $z_{\mathrm{max}}$ cannot be solved analytically; hence, we calculated $z_{\mathrm{max}}$ numerically as follows. 
 Using the luminosity $L_{\nu}$ of the source at the redshift $z$, the observed flux $f_{\nu}$ is described as follows:
\begin{equation}
\label{f_z}
f_{\nu}(z) = \frac{(1+z)^{1-\alpha}}{4\pi D_L^2(z)}L_{\nu},
\end{equation}
where $D_L$ is the luminosity distance for a given redshift in our adopted cosmology. 
 The spectral index $\alpha$ is derived from the assumption that the SED of the galaxies in the infrared obeys a simple power law $f(\nu) \propto \nu^{-\alpha}$.
 When a source is artificially moved away until the detection limit $f_{\nu}$($z_{\mathrm{max}}) = f_{\mathrm{min},\nu}$, equation (\ref{f_z}) becomes
\begin{equation}
\label{fmin_zmax}
f_{\mathrm{min},\nu} = \frac{(1+z_{\mathrm{max}})^{1-\alpha}}{4\pi D^2_{L_{\mathrm{max}}}} L_{\nu},
\end{equation}
where $D_{L_{\mathrm{max}}}$ is the maximum luminosity distance $D_L (z_{\mathrm{max}})$.
Using equations (\ref{f_z}) and (\ref{fmin_zmax}), the maximum luminosity distance can be derived as
\begin{equation}
\label{DLmax1}
D_{L_{\mathrm{max}}} = D_L \left(\frac{1+z}{1+z_{\mathrm{max}}}\right)^{\frac{\alpha -1}{2}} \left({\frac{f_{\nu}}{f_{\mathrm{min},\nu}}} \right)^{\frac{1}{2}}.
\end{equation}
On the other hand, the luminosity distance $D_L$ is defined as
\begin{equation}
\label{DL}
D_L = \frac{c}{H_0}(1+z) \int_0^z \frac{\mathrm{d}z^{\prime}}{\sqrt{\Omega_M(1+z^{\prime})^3 + \Omega_{\Lambda}}}.
\end{equation}
Hence, $D_{L_{\mathrm{max}}}$ also can be described as
\begin{equation}
\label{DLmax2}
D_{L_{\mathrm{max}}} = \frac{c}{H_0}(1+z_{\mathrm{max}}) \int_0^{z_{\mathrm{max}}} \frac{\mathrm{d}z^{\prime}}{\sqrt{\Omega_M(1+z^{\prime})^3 + \Omega_{\Lambda}}}.
\end{equation}
Therefore, we numerically estimate $z_{\mathrm{max}}$ by substituting $z$ into equations (\ref{DLmax1}) and (\ref{DLmax2}) at intervals of $\Delta z$ and iterating the above approach until the difference between the $D_{L_{\mathrm{max}}}$ values obtained from equations (\ref{DLmax1}) and (\ref{DLmax2}) is minimized. In this study, we used $\Delta z$ = $10^{-5}$.

We note that the spectral index $\alpha$ is calculated using the 9 $\micron$ and 18 $\micron$ fluxes as 
\begin{equation}
\label{alpha}
\alpha  =  -\frac{\log \left(\frac{f_{18}}{f_9} \right)}{\log \left( \frac{\nu_{18}}{\nu_9} \right)}.
\end{equation}
Although the distribution of $\alpha$ has large scatter, the average value $< \alpha >$ is $1.08 \pm 0.68$ for all samples.
 The average $< \alpha >$ slopes of Type 1s, Type 2s, and Composites are relatively large ($\geq$ 1).
 Conversely, $< \alpha >$ for SFs is relatively small ($\leq$ 1).
 LINERs have an intermediate value ($\sim$ 1). 
 In this study, we calculate $< \alpha >$ for each type of galaxy, and when an object is missing $f_{18}$ or $f_9$ information, we adopt these values, taking into account its type.

Next, we determine the completeness correction function $C(z)$. 
 First, we construct the dependence of completeness $C(f)$ for flux $f$ using the result of the log N-log S test described by \citet{Kataza}.
 Second, we convert the flux to the redshift for each object using equation (\ref{f_z}). 
 Here $L_\nu$ is treated as a constant for each object, because the luminosity of an object does not change when an object is artificially moved away with redshift.
\begin{figure}
   \begin{center}
      \FigureFile(80mm,50mm){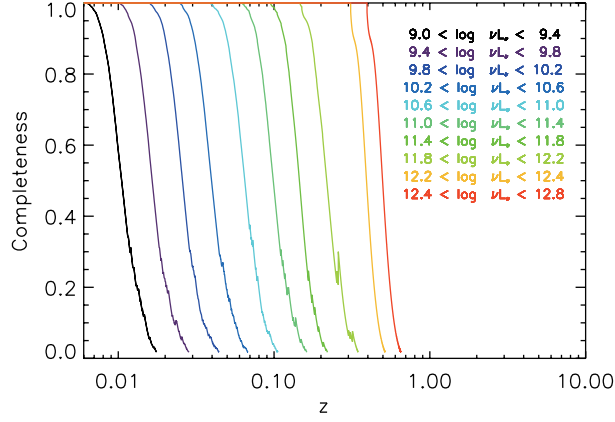}
   \end{center}
\caption{Averaged completeness as a function of redshift for 18 $\micron$ galaxies.}
\label{z_comp}
\end{figure}
Figure \ref{z_comp} shows the averaged completeness as a function of the redshift for each 18 $\micron$  luminosity bin.

\begin{figure}
  \begin{center}
    \FigureFile(80mm,50mm){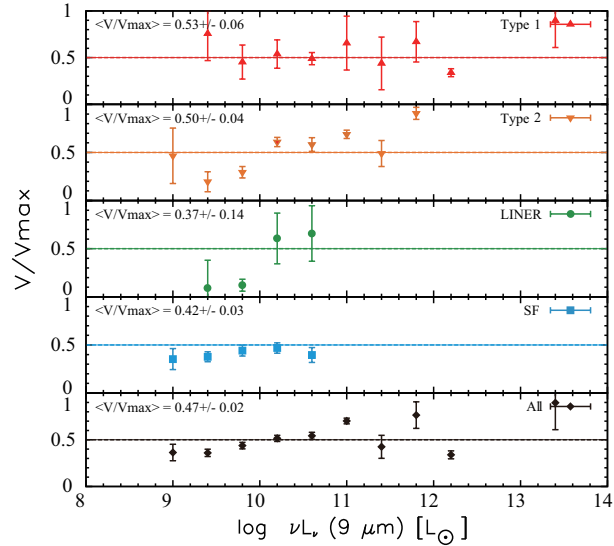}
   \end{center}   
\caption{Average values of $V/V_{\mathrm{max}}$ for the different types of galaxies as a function of 9 $\micron$ luminosity.}
\label{V_Vmax_test_09}
\end{figure}

\begin{figure}
  \begin{center}
    \FigureFile(80mm,50mm){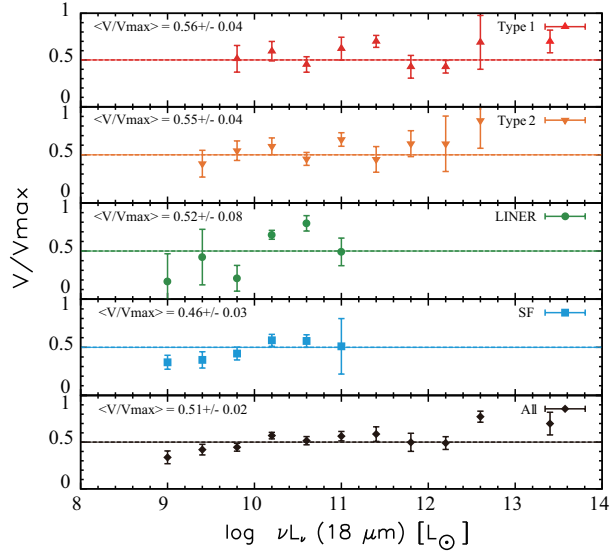}
   \end{center}   
\caption{Average values of $V/V_{\mathrm{max}}$ for the different types of galaxies as a function of 18 $\micron$ luminosity.}
\label{V_Vmax_test_18}
\end{figure}

Figures \ref{V_Vmax_test_09} and \ref{V_Vmax_test_18} show the results of the $V/V_{\mathrm{max}}$ tests after correcting for incompleteness. 
 The average value of $V/V_{\mathrm{max}}$, $<V/V_{\mathrm{max}}>$, for the entire sample at 9 $\micron$ is 0.41 $\pm$ 0.02 (before completeness correction) and 0.47 $\pm$ 0.02 (after completeness correction). 
 Similarly for 18 $\micron$, $<V/V_{\mathrm{max}}>$ is 0.46 $\pm$ 0.02 (before completeness correction) and 0.51 $\pm$ 0.02 (after completeness correction). 
 In this study, we basically assume that our samples do not show any evolution (i.e, the spatial distribution of samples does not change with redshift), because they are located in the nearby ($z\sim$ 0.04) universe. 
 Therefore, these results indicate that completeness correction works satisfactorily.

\section{The Luminosity Functions at 9 and 18 $\micron$}
\label{result}
\subsection{The Luminosity Function of Mid-Infrared Galaxies}
Here we obtain the LFs (i.e., the volume density of galaxies per unit magnitude range) of our AKARI MIR galaxies. 
The LFs are derived by following the 1/$V_{\mathrm{max}}$ method as described by \citet{Schmidt}.
The volume density $\phi (L)$ and its uncertainty $\sigma_{\phi (L)}$ are derived using the following expressions:

\begin{equation}
\phi(L) = \sum_i^N \frac{1}{V_{\mathrm{max},i}},
\end{equation}

\begin{equation}
\sigma_{\phi(L)} =\sqrt{\sum_i^N \frac{1}{V_{\mathrm{max},i}^2}},
\end{equation}
where $V_{\mathrm{max}}$ is the maximum comoving volume of $i$th galaxies after applying K-correction and completeness correction, and the sum is over all galaxies in each luminosity bin.

\begin{figure*}
	\begin{center}
    	\FigureFile(160mm,50mm){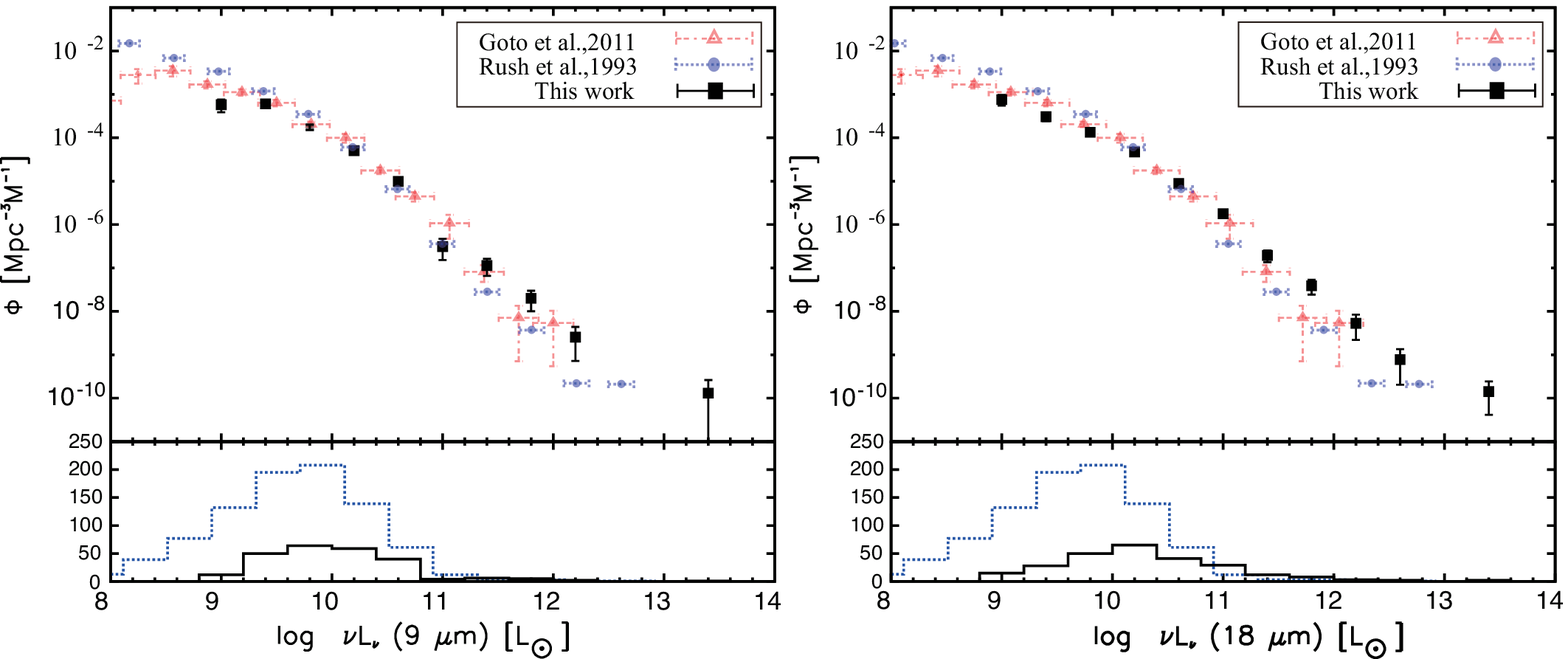} 
   	\end{center}   
	\caption{LFs at 9 (left) and 18 (right) $\micron$ for all galaxies, plotted in terms of volume density as a function of luminosity. The 12 $\micron$ LFs from Rush, Malkan,\& Spinoglio (1993) after converting $\nu L_\nu (12\, \micron)$ to $\nu L_\nu (9, 18\, \micron)$ and the TIR LFs \citep{Goto} after converting $\nu L_{TIR}$ to $\nu L_\nu (9, 18\, \micron)$ are also plotted for comparison. The vertical error bars are calculated from the Poisson statistical uncertainty and the horizontal error bars represent the uncertainty of conversion to $\nu L_\nu (9, 18\, \micron)$.}
    \label{LF}
\end{figure*}

The LFs of all galaxies at 9 and 18 $\micron$ are given in table \ref{LF_table}. 
 Figure \ref{LF} shows the resulting LFs at 9 and 18 $\micron$. 
 We also fit the LFs for all galaxies using the double-power law \citep{Marshall}:
\begin{equation}
\phi(L)\mathrm{d}L = \phi^* \left\{ \left( \frac{L}{L^*} \right)^{-\alpha} +  \left( \frac{L}{L^*} \right)^{-\beta} \right\}^{-1} \frac{\mathrm{d}L}{L^*}.
\end{equation}
 The free parameters are the characteristic luminosity $L^*$, the normalization factor $\phi^*$, the faint end slope $\alpha$, and the bright end slope $\beta$, respectively. 
 The best-fitting values are summarized in table \ref{best-fitt_LF}. 
 As shown in table \ref{best-fitt_LF}, the shapes of the 9 and 18 $\micron$ LFs are consistent with each other.

 \begin{table*}
 \caption{LF of all galaxies at 9 and 18 $\micron$.}
  \begin{center}
    \begin{tabular}{rrrrrrrr}
\hline
 & \multicolumn{3}{c}{9 $\micron$} &  & \multicolumn{3}{c}{18 $\micron$}\\
 \cline{2-4}  \cline{6-8}
$\log L$ & \multicolumn{1}{c}{$\phi$} & \multicolumn{1}{c}{1$\sigma$ err} & N &
      & \multicolumn{1}{c}{$\phi$} & \multicolumn{1}{c}{1$\sigma$ err} & N \\
 \cline{2-4}  \cline{6-8}
[$\LO$] & \multicolumn{1}{c}{[Mpc$^{-3}$ Mag$^{-1}$]} & \multicolumn{1}{c}{[Mpc$^{-3}$ Mag$^{-1}$]} &  &
& \multicolumn{1}{c}{[Mpc$^{-3}$ Mag$^{-1}$]} & \multicolumn{1}{c}{[Mpc$^{-3}$ Mag$^{-1}$]}  \\  
\hline
	9.00 &	$5.76 \times 10^{-4}$ &  $1.89 \times 10^{-4}$ & 12 & &
			$7.61 \times 10^{-4}$ &  $2.09 \times 10^{-4}$ & 15 \\   
	9.40 &  $6.06 \times 10^{-4}$ &  $9.37 \times 10^{-5}$ & 50 & &
			$3.07 \times 10^{-4}$ &  $6.35 \times 10^{-5}$ & 28 \\
	9.80 & 	$1.76 \times 10^{-4}$ &  $2.35 \times 10^{-5}$ & 64 & &
			$1.34 \times 10^{-4}$ &  $2.07 \times 10^{-5}$ & 50 \\
	10.2 & 	$5.07 \times 10^{-5}$ &  $6.94 \times 10^{-6}$ & 59 & &
			$4.72 \times 10^{-5}$ &  $6.29 \times 10^{-6}$ & 65	\\	
	10.6 &	$9.85 \times 10^{-6}$ &  $1.64 \times 10^{-6}$ & 40 & &
			$8.81 \times 10^{-6}$ &  $1.44 \times 10^{-6}$ & 41	\\	
	11.0 & 	$3.11 \times 10^{-7}$ &  $1.58 \times 10^{-7}$ &  4 & &
			$1.77 \times 10^{-6}$ &  $3.45 \times 10^{-7}$ & 29 \\
	11.4 &	$1.14 \times 10^{-7}$ &  $4.84 \times 10^{-8}$ &  6 & &
			$1.95 \times 10^{-7}$ &  $5.88 \times 10^{-8}$ & 12 \\  
	11.8 &	$1.99 \times 10^{-8}$ &  $9.84 \times 10^{-9}$ &  5 & &
			$3.88 \times 10^{-8}$ &  $1.45 \times 10^{-8}$ &  8 \\
	12.2 & 	$2.56 \times 10^{-9}$ &  $1.84 \times 10^{-9}$ &  2 & &
			$5.31 \times 10^{-9}$ &  $3.11 \times 10^{-9}$ &  3 \\	
	12.6 &	\multicolumn{1}{c}{---} & \multicolumn{1}{c}{---} & \multicolumn{1}{c}{---}  & &
			$7.74 \times 10^{-10}$ & $5.70 \times 10^{-10}$ & 2 \\ 
	13.4 &  $1.30 \times 10^{-10}$ & $1.30 \times 10^{-10}$ & 1 & &
			$1.42 \times 10^{-10}$ & $1.01 \times 10^{-10}$ & 2 \\		
\hline
    \end{tabular}
  \end{center}
    \label{LF_table}
\end{table*}

\begin{table*}
  \caption{Best double-power law fit parameters for the 9 and 18 $\micron$ LFs}
  \label{best-fitt_LF}
  \begin{center}
    \begin{tabular}{ccccc}
      \hline
\multicolumn{1}{c}{band} & \multicolumn{1}{c}{$\phi^*$ (Mpc$^{-3}$Mag$^{-1}$}) & \multicolumn{1}{c}{L$^*$ ($\LO$)}  & $\alpha$ (faint-end) & $\beta$ (bright-end) \\
\hline
 9 $\micron$ & ($1.7 \pm 0.9) \times 10^{-4}$ & $(1.2 \pm 0.4) \times 10^{10}$ & $-0.6 \pm 0.2$ & $ -2.6 \pm 0.2$ \\ 
 18 $\micron$ & $(1.0 \pm 0.8) \times 10^{-4}$ & $(1.4 \pm 0.7) \times 10^{10}$ & $-0.7 \pm 0.2$ & $ -2.1 \pm 0.2$ \\ 
     \hline
    \end{tabular}
  \end{center}
\end{table*}

\begin{figure*}
  \begin{center}
    \FigureFile(160mm,50mm){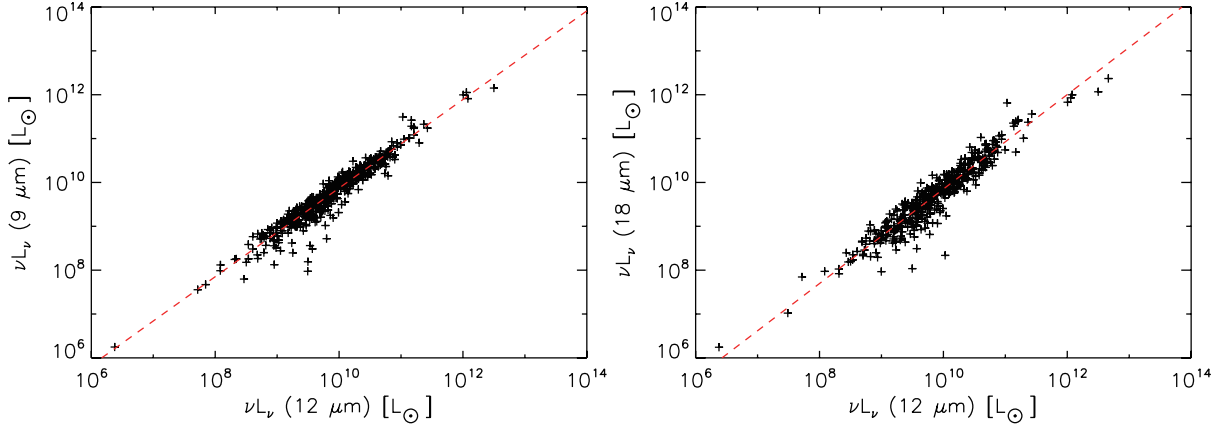}
   \end{center}   
\caption{AKARI 9 $\micron$ versus IRAS 12 $\micron$ luminosities (left) and AKARI 18 $\micron$ versus IRAS 12 $\micron$ luminosities (right). The red dotted lines show the best-fitting linear function.}
\label{nuLnu_conv}
\end{figure*}

In figure \ref{LF}, the 12 $\micron$ LFs of Rush, Malkan,\& Spinoglio (1993) and the 9 and 18 $\micron$ LFs of \citet{Goto} are also plotted for comparison, after the conversion of their wavelength to ours.
 \citet{Goto} constructed the total infrared (TIR) LFs using AKARI's six IR bands and reported the relation between $L_{TIR}$ and $\nu L_\nu (9, 18\, \micron)$.
 Therefore, we converted $L_{TIR}$ into $\nu L_\nu (9, 18\, \micron)$ using the following formulae:

\begin{eqnarray}
\log[\nu L_\nu (9\, \micron)] 		& = & (1.04 \pm 0.01) \times \log(L_{TIR}) - (1.28 \pm 0.31), \\
\log[\nu L_\nu (18\, \micron)]  	& = & (1.10 \pm 0.01) \times \log(L_{TIR}) + (1.96 \pm 0.40),
\end{eqnarray}  
as calculated by Goto \etal (2011, private communication). 
 For the 12 $\micron$ LFs, we calculated the conversion factors by plotting $\nu L_\nu (12\, \micron)$ versus $\nu L_\nu (9, 18\, \micron)$, as shown in figure \ref{nuLnu_conv} and by \citet{Goto}, and obtained these formulae:

\begin{eqnarray}
\log[\nu L_\nu (9\, \micron)] 	& = & (1.010 \pm 0.007) \times \log[\nu L_\nu (12\, \micron)] - (0.23 \pm 0.07),\\
\log[\nu L_\nu (18\, \micron)] & = & (1.077 \pm 0.007) \times \log[\nu L_\nu (12\, \micron)] - (0.92 \pm 0.07). 
\end{eqnarray}  
The shapes of the LFs obtained from previous studies (\cite{Rush}, \cite{Goto}) are in good agreement with our determinations, within the error, except for low-luminosity populations.
 The cause of the differences in the low-luminosity regime is possibly the effect of local inhomogeneities (particularly the Virgo supercluster) in the IRAS survey, as suggested by many authors. 
 For instance, \citet{Fang} selected a sample of 668 galaxies from the IRAS Faint Source Survey and derived 12 $\micron$ LFs. 
 They compared their LFs with Rush, Malkan,\& Spinoglio (1993) and explained the overestimates of number densities in the low-luminosity region as being introduced by the peculiar motion of the local supercluster. 
 The threshold redshift values adopted by Rush, Malkan ,\& Spinoglio (1993) are $1.0 \times 10^{-5}$ and $4.0 \times 10^{-4}$ by \citet{Goto}. 
 These values are much smaller than the value adopted by us in this study (0.006). 
 As a result, their LFs are probably affected in some way by the local supercluster.

\begin{figure*}
  \begin{center}
    \FigureFile(160mm,50mm){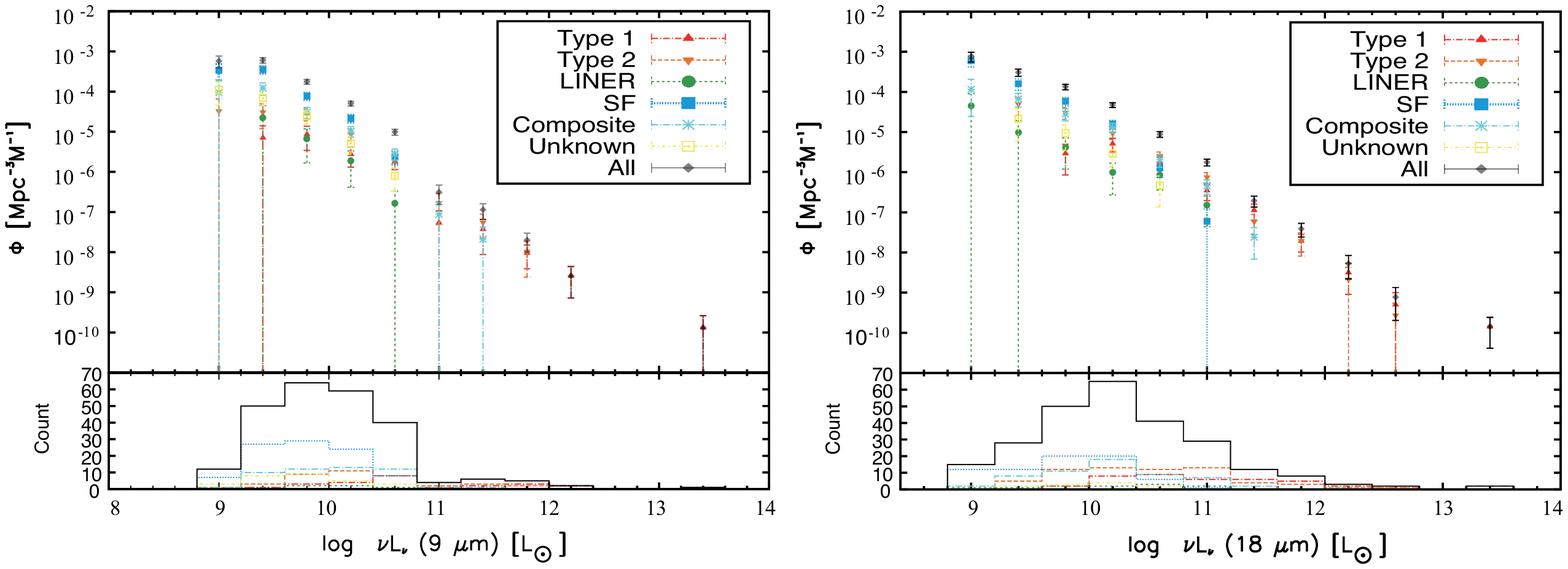}
   \end{center}   
\caption{LFs at 9 (left) and 18 (right) $\micron$ for each type of galaxy, plotted in terms of volume density as a function of luminosity. The error bars are calculated from the Poisson statistical uncertainty.}
\label{LF_type}
\end{figure*}

The LFs for each spectroscopic type of galaxies at 9 and 18 $\micron$ are given in tables \ref{LF_table_09} and \ref{LF_table_18}, respectively.
 Figure \ref{LF_type} shows the LFs at 9 and 18 $\micron$ for each type of galaxy.
 SFs are the most numerous objects at low luminosities, while AGNs dominate the volume density at luminosities above $\sim10^{11} \LO$.
 Above $\sim10^{11} \LO$, the volume densities of Type 1s and Type 2s are similar, except that Type 1s extend to a higher peak luminosity ($\gtrsim 10^{12} \LO$).
 This tendency was also reported by Rush, Malkan, \& Spmoglio (1993), based on their 12 $\micron$ flux-limited sample using the IRAS Faint Source Catalog (FSC).

 \begin{table*}
 \caption{9 $\micron$ LF for various spectroscopic galaxies.}
  \begin{center}
    \begin{tabular}{rrrrrrrr}
\hline
 & \multicolumn{3}{c}{Type 1 AGN} &  & \multicolumn{3}{c}{Type 2 AGN}\\
 \cline{2-4}  \cline{6-8}
$\log L$ & \multicolumn{1}{c}{$\phi$} & \multicolumn{1}{c}{1$\sigma$ err} & N &
      & \multicolumn{1}{c}{$\phi$} & \multicolumn{1}{c}{1$\sigma$ err} & N \\
 \cline{2-4}  \cline{6-8}
[$\LO$] & \multicolumn{1}{c}{[Mpc$^{-3}$ Mag$^{-1}$]} & \multicolumn{1}{c}{[Mpc$^{-3}$ Mag$^{-1}$]} &  &
& \multicolumn{1}{c}{[Mpc$^{-3}$ Mag$^{-1}$]} & \multicolumn{1}{c}{[Mpc$^{-3}$ Mag$^{-1}$]}  \\  
\hline
	9.00 &       \multicolumn{1}{c}{---} &       \multicolumn{1}{c}{---} &  & &  $3.28\times 10^{-5}$ &  $3.28\times 10^{-5}$ & 1 \\ 
	9.40 &  $7.03\times 10^{-6}$ &  $7.03\times 10^{-6}$ & 1 & &  $3.06\times 10^{-5}$ &  $1.87\times 10^{-5}$ & 3 \\
	9.80 &  $8.54\times 10^{-6}$ &  $5.12\times 10^{-6}$ & 3 & &  $2.83\times 10^{-5}$ &  $9.64\times 10^{-6}$ & 9 \\
	10.2 &  $2.73\times 10^{-6}$ &  $1.41\times 10^{-6}$ & 4 & &  $8.85\times 10^{-6}$ &  $2.79\times 10^{-6}$ & 11 \\
	10.6 &  $1.81\times 10^{-6}$ &  $6.80\times 10^{-7}$ & 8 & &  $1.86\times 10^{-6}$ &  $6.88\times 10^{-7}$ & 8 \\
	11.0 &  $5.37\times 10^{-8}$ &  $5.37\times 10^{-8}$ & 1 & &  $1.69\times 10^{-7}$ &  $1.20\times 10^{-7}$ & 2 \\
	11.4 &  $3.71\times 10^{-8}$ &  $2.84\times 10^{-8}$ & 2 & &  $5.57\times 10^{-8}$ &  $3.30\times 10^{-8}$ & 3 \\
	11.8 &  $9.47\times 10^{-9}$ &  $5.62\times 10^{-9}$ & 3 & &  $1.05\times 10^{-8}$ &  $8.07\times 10^{-9}$ & 2 \\
	12.2 &  $2.56\times 10^{-9}$ &  $1.84\times 10^{-9}$ & 2 & &       \multicolumn{1}{c}{---} &  \multicolumn{1}{c}{---} &  \\	
	13.4 &  $1.30\times 10^{-10}$ &  $1.32\times 10^{-10}$ & 1 & &       \multicolumn{1}{c}{---} &       \multicolumn{1}{c}{---} & \\		
\hline
    \end{tabular}
        \begin{tabular}{rrrrrrrr}
\hline
 & \multicolumn{3}{c}{LINER} &  & \multicolumn{3}{c}{Star-Forming}\\
 \cline{2-4}  \cline{6-8}
$\log L$ & \multicolumn{1}{c}{$\phi$} & \multicolumn{1}{c}{1$\sigma$ err} & N &
      & \multicolumn{1}{c}{$\phi$} & \multicolumn{1}{c}{1$\sigma$ err} & N \\
 \cline{2-4}  \cline{6-8}
[$\LO$] & \multicolumn{1}{c}{[Mpc$^{-3}$ Mag$^{-1}$]} & \multicolumn{1}{c}{[Mpc$^{-3}$ Mag$^{-1}$]} &  &
& \multicolumn{1}{c}{[Mpc$^{-3}$ Mag$^{-1}$]} & \multicolumn{1}{c}{[Mpc$^{-3}$ Mag$^{-1}$]}    \\
	\hline
	9.00 &       \multicolumn{1}{c}{---} &       \multicolumn{1}{c}{---} & & &  $3.39 \times 10^{-4}$&  $1.47 \times 10^{-4}$ & 7 \\
	9.40 & 	$2.23\times 10^{-5}$ &  $2.23\times 10^{-5}$ & 1 & &  $3.54 \times 10^{-4}$ &  $7.42\times 10^{-5}$ &27 \\
	9.80 & $6.71\times 10^{-6}$ &  $5.05\times 10^{-6}$ & 2 & &  $7.74\times 10^{-5}$ &  $1.53\times 10^{-5}$ &29 \\
	10.2 &  $1.87\times 10^{-6}$ &  $1.46\times 10^{-6}$ & 2 & &  $2.17\times 10^{-5}$ &  $4.61\times 10^{-6}$ &24 \\
	10.6 & $1.66\times 10^{-7}$ &  $1.66\times 10^{-7}$ & 1 & &  $2.38\times 10^{-6}$ &  $8.68\times 10^{-7}$ & 8 \\
\hline
    \end{tabular}  
            \begin{tabular}{rrrrrrrr}
\hline
 & \multicolumn{3}{c}{Composite} &  & \multicolumn{3}{c}{Unknown}\\
 \cline{2-4}  \cline{6-8}
$\log L$ & \multicolumn{1}{c}{$\phi$} & \multicolumn{1}{c}{1$\sigma$ err} & N &
      & \multicolumn{1}{c}{$\phi$} & \multicolumn{1}{c}{1$\sigma$ err} & N \\
 \cline{2-4}  \cline{6-8}
[$\LO$] & \multicolumn{1}{c}{[Mpc$^{-3}$ Mag$^{-1}$]} & \multicolumn{1}{c}{[Mpc$^{-3}$ Mag$^{-1}$]} &  &
& \multicolumn{1}{c}{[Mpc$^{-3}$ Mag$^{-1}$]} & \multicolumn{1}{c}{[Mpc$^{-3}$ Mag$^{-1}$]}    \\
	\hline
	      9.00 &  9.43$\times 10^{-5}$ &  9.43$\times 10^{-5}$ &    1 & &  1.10$\times 10^{-4}$ &  6.43$\times 10^{-5}$ &    3 \\
      9.40 &  1.23$\times 10^{-4}$ &  4.18$\times 10^{-5}$ &   10 & &  6.90$\times 10^{-5}$ &  2.50$\times 10^{-5}$ &    8 \\
      9.80 &  3.09$\times 10^{-5}$ &  9.63$\times 10^{-6}$ &   12 & &  2.44$\times 10^{-5}$ &  8.92$\times 10^{-6}$ &    9 \\
      10.2 &  1.05$\times 10^{-5}$ &  3.05$\times 10^{-6}$ &   13 & &  5.13$\times 10^{-6}$ &  2.39$\times 10^{-6}$ &    5 \\
      10.6 &  2.84$\times 10^{-6}$ &  8.67$\times 10^{-7}$ &   12 & &  7.93$\times 10^{-7}$ &  4.59$\times 10^{-7}$ &    3 \\
      11.0 &  8.84$\times 10^{-8}$ &  8.84$\times 10^{-8}$ &    1 & &       \multicolumn{1}{c}{---} &       \multicolumn{1}{c}{---} &  \\
      11.4 &  2.11$\times 10^{-8}$ &  2.11$\times 10^{-8}$ &    1 & &       \multicolumn{1}{c}{---} &       \multicolumn{1}{c}{---} &  \\
	\hline
    \end{tabular}    
  \end{center}
  \label{LF_table_09}
\end{table*}

\begin{table*}
	\caption{18 $\micron$ LF for various spectroscopic galaxies.}
	\begin{center}
    	\begin{tabular}{rrrrrrrr}
		\hline
 			  & \multicolumn{3}{c}{Type 1 AGN} &  & \multicolumn{3}{c}{Type 2 AGN}\\
 \cline{2-4}  \cline{6-8}
		$\log L$ & \multicolumn{1}{c}{$\phi$} & \multicolumn{1}{c}{1$\sigma$ err} & N & & \multicolumn{1}{c}{$\phi$} & \multicolumn{1}{c}{1$\sigma$ err} & N \\
 \cline{2-4}  \cline{6-8}
		[$\LO$] & \multicolumn{1}{c}{[Mpc$^{-3}$ Mag$^{-1}$]} & \multicolumn{1}{c}{[Mpc$^{-3}$ Mag$^{-1}$]} &  &
& \multicolumn{1}{c}{[Mpc$^{-3}$ Mag$^{-1}$]} & \multicolumn{1}{c}{[Mpc$^{-3}$ Mag$^{-1}$]}  \\  
\hline
	9.40 &       \multicolumn{1}{c}{---} &       \multicolumn{1}{c}{---} & & &  $4.99\times 10^{-5}$ &  $2.41\times 10^{-5}$ & 5 \\
	9.80 &  $2.92\times 10^{-6}$ &  $2.07\times 10^{-6}$ & 2 & &  $3.06\times 10^{-5}$ &  $9.84\times 10^{-6}$ & 12 \\
	10.2 &  $5.10\times 10^{-6}$ &  $1.91\times 10^{-6}$ & 8 & &  $9.32\times 10^{-6}$ &  $2.71\times 10^{-6}$ & 13 \\
	10.6 &  $1.71\times 10^{-6}$ &  $6.08\times 10^{-7}$ & 9 & &  $2.45\times 10^{-6}$ &  $7.34\times 10^{-7}$ & 12 \\
	11.0 &  $3.48\times 10^{-7}$ &  $1.53\times 10^{-7}$ & 6 & &  $7.52\times 10^{-7}$ &  $2.18\times 10^{-7}$ & 13 \\
	 11.4 &  $1.12\times 10^{-7}$ &  $4.76\times 10^{-8}$ & 6 & &  $5.86\times 10^{-8}$ &  $2.98\times 10^{-8}$ & 4 \\
	 11.8 &  $1.93\times 10^{-8}$ &  $9.06\times 10^{-9}$ & 5 & &  $1.95\times 10^{-8}$ &  $1.13\times 10^{-8}$ & 3 \\
	 12.2 &  $3.17\times 10^{-9}$ &  $2.26\times 10^{-9}$ & 2 & &  $2.13\times 10^{-9}$ &  $2.13\times 10^{-9}$ & 1 \\
	 12.6 &  $5.01\times 10^{-10}$ &  $5.01\times 10^{-10}$ & 1 & &  $2.72\times 10^{-10}$ &  $2.72\times 10^{-10}$ & 1 \\
	 13.4 &  $1.42\times 10^{-10}$ &  $1.01\times 10^{-10}$ & 2 & &       \multicolumn{1}{c}{---} &  \multicolumn{1}{c}{---} &  \\
	 \hline
	    \end{tabular} 	    
         \begin{tabular}{rrrrrrrr}
\hline
 & \multicolumn{3}{c}{LINER} &  & \multicolumn{3}{c}{Star-Forming}\\
 \cline{2-4}  \cline{6-8}
$\log L$ & \multicolumn{1}{c}{$\phi$} & \multicolumn{1}{c}{1$\sigma$ err} & N &
      & \multicolumn{1}{c}{$\phi$} & \multicolumn{1}{c}{1$\sigma$ err} & N \\
 \cline{2-4}  \cline{6-8}
[$\LO$] & \multicolumn{1}{c}{[Mpc$^{-3}$ Mag$^{-1}$]} & \multicolumn{1}{c}{[Mpc$^{-3}$ Mag$^{-1}$]} &  &
& \multicolumn{1}{c}{[Mpc$^{-3}$ Mag$^{-1}$]} & \multicolumn{1}{c}{[Mpc$^{-3}$ Mag$^{-1}$]}    \\
	\hline
	9.00 &  $4.48\times 10^{-5}$ &  $4.48\times 10^{-5}$ & 1 & &  $6.00 \times 10^{-4}$ &  $1.82 \times 10^{-4}$ & 12 \\
	9.40 & 	$9.68\times 10^{-6}$ &  $9.68\times 10^{-6}$ & 1 & &  $1.59 \times 10^{-4}$ &  $5.00\times 10^{-5}$ & 12 \\
	9.80 & $4.17\times 10^{-6}$ &  $2.98\times 10^{-6}$ & 2 & &  $5.83\times 10^{-5}$ &  $1.39\times 10^{-5}$ & 20 \\
	10.2 &  $9.80\times 10^{-7}$ & $7.08\times 10^{-7}$ & 2 &  & $1.58\times 10^{-5}$ &  $3.76 \times 10^{-6}$ & 20 \\
	10.6 & $8.55\times 10^{-7}$ &  $4.97\times 10^{-7}$ & 3 & &  $1.27\times 10^{-6}$ &  $5.36\times 10^{-7}$ & 6 \\
	11.0 & $1.51\times 10^{-7}$ &   $1.06\times 10^{-7}$ & 2 & &  $6.00\times 10^{-8}$ &  $6.00\times 10^{-8}$ & 1 \\
\hline
    \end{tabular}    
         \begin{tabular}{rrrrrrrr}
\hline
 & \multicolumn{3}{c}{Composite} &  & \multicolumn{3}{c}{Unknown}\\
 \cline{2-4}  \cline{6-8}
$\log L$ & \multicolumn{1}{c}{$\phi$} & \multicolumn{1}{c}{1$\sigma$ err} & N &
      & \multicolumn{1}{c}{$\phi$} & \multicolumn{1}{c}{1$\sigma$ err} & N \\
 \cline{2-4}  \cline{6-8}
[$\LO$] & \multicolumn{1}{c}{[Mpc$^{-3}$ Mag$^{-1}$]} & \multicolumn{1}{c}{[Mpc$^{-3}$ Mag$^{-1}$]} &  &
& \multicolumn{1}{c}{[Mpc$^{-3}$ Mag$^{-1}$]} & \multicolumn{1}{c}{[Mpc$^{-3}$ Mag$^{-1}$]}    \\
	\hline
      9.00 &  1.16$\times 10^{-4}$ &  9.11$\times 10^{-5}$ &    2 & &       \multicolumn{1}{c}{---} &       \multicolumn{1}{c}{---} &  \\
      9.40 &  6.67$\times 10^{-5}$ &  2.48$\times 10^{-5}$ &    8 & &  2.15$\times 10^{-5}$ &  1.57$\times 10^{-5}$ &    2 \\
      9.80 &  2.84$\times 10^{-5}$ &  9.47$\times 10^{-6}$ &   11 & &  9.41$\times 10^{-6}$ &  5.94$\times 10^{-6}$ &    3 \\
      10.2 &  1.30$\times 10^{-5}$ &  3.37$\times 10^{-6}$ &   18 & &  2.91$\times 10^{-6}$ &  1.58$\times 10^{-6}$ &    4 \\
      10.6 &  2.06$\times 10^{-6}$ &  7.26$\times 10^{-7}$ &    9 & &  4.70$\times 10^{-7}$ &  3.34$\times 10^{-7}$ &    2 \\
      11.0 &  4.64$\times 10^{-7}$ &  1.82$\times 10^{-7}$ &    7 & &       \multicolumn{1}{c}{---} &       \multicolumn{1}{c}{---} &  \\
      11.4 &  2.42$\times 10^{-8}$ &  1.73$\times 10^{-8}$ &    2 & &       \multicolumn{1}{c}{---} &       \multicolumn{1}{c}{---} &  \\
	 \hline
    \end{tabular}  
  \end{center}
    \label{LF_table_18}
\end{table*}
 
\section{DISCUSSION}

\subsection{Number Density Ratio of Type 1 AGNs and Type 2 AGNs}
We compare the number density of Type 1 AGNs with that of Type 2s to constrain the structure of the hypothesized torus invoked by unification. 
 We assume that the MIR luminosity of Type 1s and Type 2s should be intrinsically the same. 
 In this study, we use 18 $\micron$ luminosity, which is expected to be direct radiation from the dust torus, uninfluenced by dust extinction (the 9 $\micron$ flux, in contrast, may be more affected by silicate absorption or emission). 
 By integrating the LFs of Type 1s and Type 2s separately, we obtain the number density ratio, $\Phi_{\mathrm{Type\, 2}}/\Phi_{\mathrm{Type\, 1}}$. 
 Here the number density $\Phi$ is given by the formula
\begin{equation}
\Phi = \int_L \phi(L) \mathrm{d}L \sim \sum_i \phi_i(L)  \Delta L,
\end{equation}
where the integral range is $\log (\nu L_{\nu})$ $>$ $10^{10}$ $L_{\odot}$ for both AGN types. 
 The resulting number density ratio of Type 2s to Type 1s, $\Phi_{\mathrm{Type\, 2}}/\Phi_{\mathrm{Type\, 1}}$, is 1.73 $\pm$ 0.36. 
 We then compared this value with previous results obtained from optical LFs \citep{Haob}, IR LFs \citep{Rush}, and hard X-ray LFs \citep{Burlon}. 
 Note that before comparing these results with ours, we converted each $L_{\nu}$ into $\nu L_\nu (18\,  \micron)$ using the SED template \citep{Shang} as follows:
 \begin{eqnarray}
\log(L_{\mathrm{[OIII]}}) 				& = & \log[\nu L_\nu (18\, \micron)] - 2.47, \\
\log[\nu L_\nu (12\, \micron)] 			& = & \log[\nu L_\nu (18\, \micron)] - 0.04, \\
\log(L_{\mathrm{15 keV}}) 				& = & \log[\nu L_\nu (18\, \micron)] - 0.61. 
\end{eqnarray}   
Table \ref{Comp_Phi} shows the results of the comparison after applying this luminosity conversion.
 
\begin{table}
\caption{Comparison of the number density ratio of Type 2 to Type 1 AGNs ($\Phi_{\mathrm{Type\, 2}}/\Phi_{\mathrm{Type\, 1}}$) with some previous studies.}
\label{Comp_Phi}
 	\begin{center}
    	\begin{tabular}{cccc}
    		\hline
      		$\Phi_{\mathrm{Type\, 2}}$($L \geq 10^{10} \LO$) /$\Phi_{\mathrm{Type\, 1}}$ ($L \geq 10^{10} \LO$) & wavelength & reference \\
      		\hline
 			$\sim$1.63 & 15-55 keV & \citet{Burlon} \\
 			$\sim$1.75 & 12 $\micron$ & \citet{Rush} \\
 			1.73 $\pm$ 0.36  & 18 $\micron$ & This work \\
 			$\sim$0.67  & [OIII] & \citet{Haob} \\
      		\hline
    	\end{tabular}
  	\end{center}
\end{table}

First, we compare our result with that of \citet{Haob}. 
 \citet{Haob} selected 3,000 AGNs using the SDSS redshift range of 0 $\leq z \leq$ 0.15 and presented the emission line LFs for Type 1s and Type 2s. 
 In this study, we focus on the luminosity of the [OIII] line as a tracer of AGN activity. 
 [OIII] can be excited by not only an AGN but also by massive stars in star-forming galaxies. 
 However, such galaxies cannot contaminate the AGN samples, because the BPT line ratio classification will exclude star-forming galaxies from our AGN sample because of their [NII]/H$\alpha$ ratios. 
 The Type 2/Type 1 ratio obtained from the [OIII] LF in \citet{Haob} is $\sim$0.67, which is smaller than our result (see table \ref{Comp_Phi}). 
 This difference may indicate that Hao et al. systematically underestimated the intrinsic [OIII] luminosities of Type 2s. 
 This situation may be occurred when the [OIII] emission from NLR of Type 2s are partially blocked by the dust torus. 
 However, above explanation would contradict the basic premise of simple unification that is expected the intrinsic differences do not exist between Type 1 and Type 2 AGNs.

Next, we compared our result with Rush, Malkan, \& Spinoglio (1993) and \citet{Burlon}. 
 As mentioned in section \ref{result}, Rush, Malkan, \& Spinoglio (1993) constructed 12 $\micron$ LFs for Type 1 and Type 2s using the IRAS FSC.
 \citet{Burlon} also constructed AGN samples in the local universe ($z \leq$ 0.1) using data from the Swift-BAT telescope and calculated the X-ray (15-55 keV) LFs of absorbed and unabsorbed AGNs, which were classified according to their absorbing column density ($N_H$).
 The results obtained from the 12 $\micron$ and hard X-ray LFs are $\sim$1.75 and $\sim$1.63, respectively, as shown in table \ref{Comp_Phi}, which is in good agreement with our results. 
 This indicates that there are a large number of narrow-line AGNs in the local universe. 
 This is also consistent with previous conclusions from X-ray observations (\cite{Maiolino+98}; \cite{Risaliti}; \cite{Malizia}). 
 Note that the classification by X-ray (unabsorbed and absorbed AGNs) does not necessarily correspond to optical classifications (Type 1 and Type 2s). 
 Some absorbed AGN samples in \citet{Burlon} tended to be classified as Type 1s by optical observations.
 This tendency indicates that the number density ratio $\Phi_{\mathrm{Type\, 2}}/\Phi_{\mathrm{Type\, 1}}$ obtained from X-ray classification would be slightly larger than that from optical classification, which is further consistent with our estimate.

\subsection{Implication for the Covering Factor of the Dust Torus}
\begin{figure}
	\begin{center}
    \FigureFile(80mm,50mm){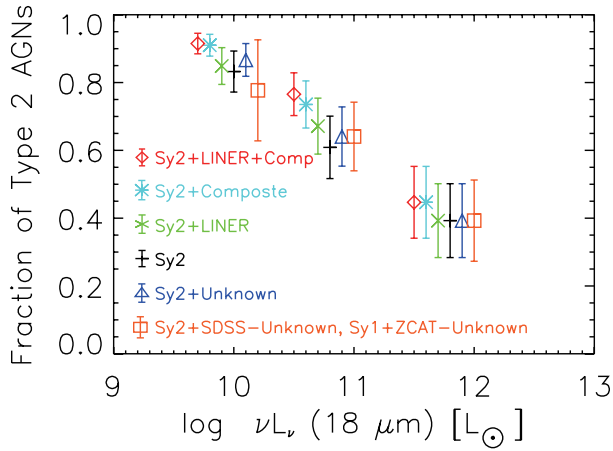}
	\end{center}   
\caption{Fraction of Type 2 AGNs to the all AGNs versus 18 $\micron$ luminosity at 0.006 $\leq$ z $\leq$ 0.2. Each symbol represents which populations were included with Type 2 or Type 1s. We basically considered only Sy2s to be Type 2s (black plus). However, we also estimated the fraction when we considered Sy2 + LINERs + Composites (red diamond), Sy2 + Composites (aqua asterisk), Sy2 + LINERs (green cross), and Sy2 + Unknown (blue triangle) as Type 2s. Furthermore, we also estimated the fraction under the assumption that the Unknown-SDSS galaxies are Type 2s and the Unknown ZCAT galaxies are Type 1s (orange square). Each symbol except for black plus (Sy2) was moved aside to avoid complexity.}
\label{N_ratio}
\end{figure}

Figure \ref{N_ratio} shows the fraction of Type 2 AGNs as a function of 18 $\micron$ luminosity. 
 It is a decreasing function. 
 Note that the redshift range of the Type 2s plotted in figure \ref{N_ratio} is limited to 0.006 $\leq z \leq$ 0.2, because the SDSS spectroscopic catalog may not be as complete for Type 2s at $z \geq$ 0.2 compared with Type 1s. 
 Before interpreting this result, we should consider the effect of (i) our classification of Type 2, (ii) unknown types of galaxies, (iii) our ``rejected'' galaxies, and (iv) the detection limit of the H$\alpha$ line for the SDSS spectroscopic survey. 

\subsubsection{Effect of Type 2 AGN classification} 
In this study, we basically only considered Seyfert 2 galaxies to be Type 2s. 
 However, most of the LINERs seem to also follow the distribution of Type 2s.
 Moreover, some of the Composites can harbor low-luminosity Type 2s. 
 Hence, we examined the effect these populations might have if they were included in the fraction of Type 2s. 
 For this, we made four possible groups of Type 2s.
 \begin{enumerate}
 \item Seyfert 2 galaxies (Sy2s)
 \item Sy2s + LINERs
 \item Sy2s + Composites
 \item Sy2s + LINERs + Composites
 \end{enumerate}
 Figure \ref{N_ratio} also shows the fraction of the Type 2s when LINERs and/or Composites were added to the Type 2 category as described above. 
 As shown in this figure, the fraction of Type 2s decreases with 18 $\micron$ luminosity regardless of the changes in the Type 2 classification criteria. 

\subsubsection{Effect of Unknown types of galaxies}  
As we described in section \ref{Classification}, 11 galaxies were classified as Unknown types of galaxies for the 18 $\micron$ sample.
 The redshift range of these galaxies is less than 0.2. 
 Among the 11 galaxies, 6 were obtained from the SDSS, and 5 were obtained from the ZCAT. 
 We found that none of these SDSS galaxies were Type 1s (see section \ref{Classification}).
 On the other hand, we could not constrain the 5 ZCAT galaxies, because the ZCAT does not include spectroscopic line information.
 Therefore, we considered the effect of these objects by including two extreme possibilities as follows:
\begin{itemize}
\item 6 SDSS galaxies and 5 ZCAT galaxies are Type 2s.
\item 6 SDSS galaxies are Type 2s and 5 ZCAT galaxies are Type 1s.
\end{itemize}
Figure \ref{N_ratio} also shows the fractions of Type 2s considering both of these cases. 
As shown in this figure, the fraction of Type 2s still decreases with 18 $\micron$ luminosity regardless of the classification of the Unknown objects.

\subsubsection{Effect of rejected galaxies}  
In section \ref{sample_selection}, we selected 243 and 255 galaxies at 9 and 18 $\micron$, respectively.
 However, it should be noted that when we cross-matched the AKARI-ZCAT objects, 1,331 AKARI sources did not match within 12 arcsec; thus, we rejected these objects for this study. 
 If these objects were galaxies, our results would be affected. 
 For an estimate of their effect, these 1,331 objects (hereinafter AKARI-non-Tycho 2 objects) were cross-matched with the 2MASS Point Source Catalog (Skrutskie et al. 2006) to investigate their color distributions. 
 We adopted 3 arcsec as our search radius, which was optimized for cross-identification between these two catalogs \citep{Ishihara}.

\begin{figure}
	\begin{center}
    \FigureFile(80mm,50mm){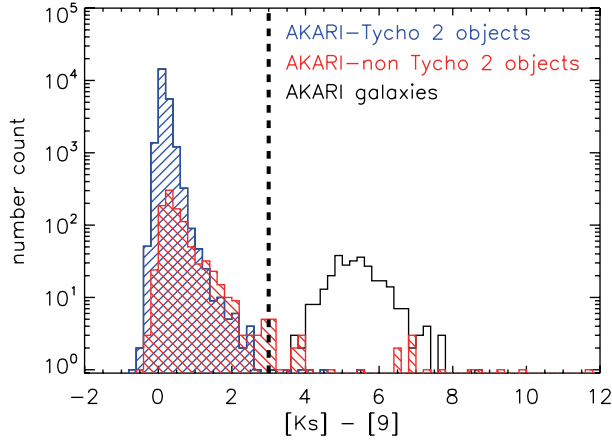}
	\end{center}   
\caption{Color ([Ks] - [9]) distribution of AKARI sources. The blue line represents the AKARI-Tycho 2 objects, the red line represents the AKARI-non-Tycho 2 objects, and the black line represents the AKARI galaxies. The dotted line indicates the threshold between stars and galaxies. We extracted the AKARI-non Tycho 2 objects with [Ks] - [9] $\geq$ 3 as galaxy candidates.}
\label{color}
\end{figure}

\begin{figure}
	\begin{center}
    \FigureFile(80mm,50mm){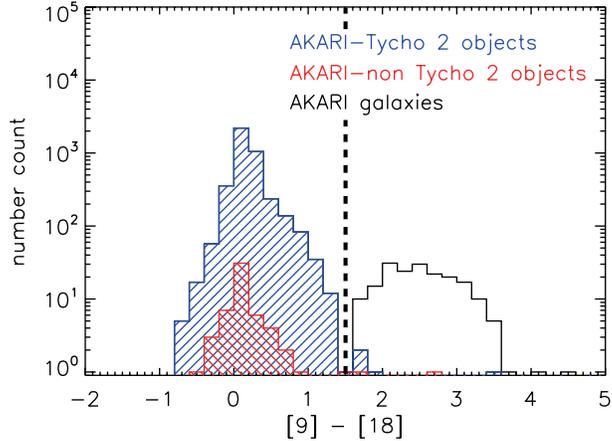}
	\end{center}   
\caption{Color ([9]-[18]) distribution of AKARI sources. The blue line represents the AKARI-Tycho 2 objects, the red line represents the AKARI-non-Tycho 2 objects, and black line represents the AKARI galaxies. The dotted line indicates the threshold between stars and galaxies. We extracted the AKARI-non-Tycho 2 objects with [9]-[18] $\geq$ 1.5 as galaxy candidates.}
\label{color2}
\end{figure}

Figures \ref{color} and \ref{color2} show the histogram of [Ks]-[9] and [9]-[18] for AKARI-non-Tycho 2 objects. 
 Here [Ks], [9], and [18] represent Vega magnitudes in the 2MASS Ks (2.2 $\micron$), AKARI 9 $\micron$, and AKARI 18 $\micron$ bands, respectively. 
 The zero point magnitudes for the 9 and 18 $\micron$ bands are 56.3 and 12.0 Jy, respectively \citep{Tanabe}.
 To ensure the reliability of the color diagram, the AKARI-Tycho 2 sources and our samples in this study (AKARI galaxies) were also plotted in these figures. 
 As shown in the figures, the AKARI-Tycho 2 stars were well-separated from the AKARI galaxies.
 Thus, color distribution distinguishes stars from galaxies. 
 In this study, we extracted galaxy candidates using these two colors. 
 We first extracted 27 AKARI-non-Tycho 2 objects that met the following criterion as galaxy candidates:
\begin{equation}
[Ks] - [9] \geq 3.0.
\end{equation}
323 AKARI-non-Tycho 2 objects could not be classified as a star or galaxy. 
 These objects are missing one or both of [Ks] and [9] data; thus, we could not investigate their [Ks] - [9] color. 
 For these 323 objects, we examined the [9]-[18] color distribution (figure \ref{color2}) and extracted 2 objects that met the following criterion as galaxy candidates:
\begin{equation}
[9] - [18] \geq 1.5.
\end{equation}
Also, 265 AKARI-non-Tycho 2 objects could not be identified as a star or galaxy, because they did not have one or both of [9] or [18] data; thus, we could not investigate their [9]-[18] color.
 As a result, 27 + 2 + 265 = 294 objects (278 objects for 9 $\micron$ and 31 objects for 18 $\micron$) remained as galaxy candidates. 
 For these objects, we carefully determined their type using the previous literature. 
 Finally, 276 objects were confirmed as stars or planetary nebulae, thus, 294 - 276 = 18 objects remained as galaxies (13 objects for 9 $\micron$ and 7 objects for 18 $\micron$).
 The maximum uncertainty caused by including them as galaxies would be 13/243 $\sim$0.053 (5.3\%) for the 9 $\micron$ sample and 7/255 $\sim$0.027 (2.7\%) for the 18 $\micron$ sample, respectively.

\subsubsection{Effect of the detection limit of the H$\alpha$ line for the SDSS spectroscopic survey}  
\begin{figure}
	\begin{center}
    \FigureFile(80mm,50mm){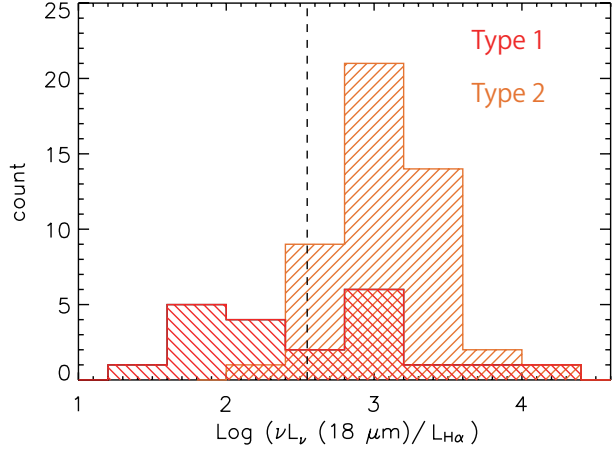}
	\end{center}   
\caption{Distribution of $\nu L_\nu (18\, \micron)$\,/\,$L_{H\alpha}$. The dashed line represents the average value of the above ratio.}
\label{L18_Ha}
\end{figure}

Finally, we consider the SDSS detection limit for the detection of the broad H$\alpha$ line from the Type 1s in the lowest luminosity bin. 
 If the detection limit was not sufficiently low to detect some Type 1s, the fraction of Type 2s could be overestimated, especially in the lowest luminosity bin in figure \ref{N_ratio}.
 To investigate this, we estimated the detection limit of the H$\alpha$ line, considering the typical luminosity ratio between H$\alpha$ and 18 $\micron$ of Type 1s. 
 Figure \ref{L18_Ha} shows the distribution of luminosity ratio of the 18 $\micron$ luminosity to the H$\alpha$ luminosity for the AKARI-SDSS AGNs.

 In this study, we adopted $\log[\nu L_\nu (18\, \micron)$/$L_{H\alpha}] \sim$2.55 as the typical luminosity ratio between the H$\alpha$ and 18 $\micron$ of Type 1s.
 Next, we investigated all AGNs in the SDSS DR7 spectroscopic region.
 According to the FWHM of H$\alpha$ and the BPT diagram, 26,568 AGNs were then extracted. 
 Furthermore, we extracted 652 AGNs with a redshift between 0.006 and 0.03，focusing on the lowest luminosity bin ($\sim 10^{10} \LO$) in figure \ref{N_ratio}, which corresponds to $z \sim$ 0.03 (see figure \ref{L}).

 \begin{figure}
	\begin{center}
    \FigureFile(80mm,50mm){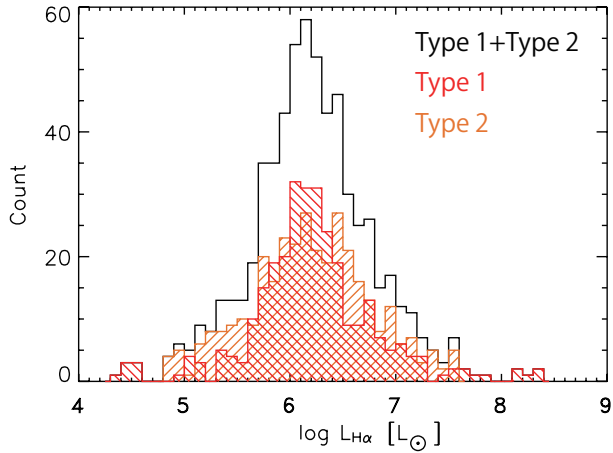}
	\end{center}   
\caption{Distribution of $L_{H\alpha}$ for all SDSS AGNs (0.006 $\leq$ z $\leq$ 0.03).}
\label{LHa}
\end{figure}
Figure \ref{LHa} shows the distribution of $L_{H\alpha}$ for the 652 AGNs. 
 As shown in figure \ref{LHa}, these histograms go down to the faintest bin of $L_{H\alpha}$ $\sim 10^5 \LO$, which is consistent with the histogram of $L_{H\alpha}$ obtained by \citep{Haob}.
 This limit corresponds to $\sim10^{7-8} \LO$ for our 18 $\micron$ AGNs, assuming the typical luminosity ratio. 
 Therefore, the detection limit seems sufficiently low to detect the broad H$\alpha$ lines from the Type 1s in the lowest luminosity bin ($\sim 10^{10} \LO$) even if we adopted the maximum value ($\sim$ 4) of $\nu L_\nu (18\, \micron)/L_{H\alpha}$ distribution as the typical luminosity ratio.

The decreasing fraction of Type 2s at increasing luminosity could mean that the luminosity from the central engine influences the structure of the torus. 
 The covering factor of the dust torus is smaller for more luminous AGNs, possibly because increasing luminosity sublimates more dust, causing the inner radius of the dust torus to recede \citep{Lawrence}. 
 This result has been reported several times in the past. 
 For example, \citet{Maiolino+07} found that the MIR spectra of 25 AGNs taken with IRS on board the Spitzer Space Telescope showed a negative correlation between $\nu L_{6.7}/\nu L_{5100}$ and the [OIII]$\lambda$5007 line luminosity. 
 $L_{6.7}$ and $L_{5100}$ are the continuum luminosity at the rest-frame wavelength of 6.7 $\micron$ and 5100 $\AA$, respectively. 
 They mentioned that one possible explanation was that higher [OIII] luminosities imply a more luminous central engine and larger dust sublimation radius, suggesting that a larger dust sublimation radius would give a lower covering factor of dust if the obscuring medium is distributed in a disk with constant height. 
 Recently, \citet{Burlon} reported a negative correlation between the fraction of absorbed AGN and hard X-ray luminosity using a Swift-BAT sample. 
 Our study confirms and extends these results with AKARI, by taking our larger MIR sample. 
 We emphasize that a luminosity-dependent torus geometry destroys the simplicity of the original torus unification scheme, and now requires that at least one new free function must be determined. 
 The torus covering fraction as a function of AGN luminosity can be measured from complete MIR observations, with our figure \ref{N_ratio} as a first effort to do so.

\section{SUMMARY}
Using the AKARI MIR all-sky survey, we constructed the 9 and 18 $\micron$ LFs for all types of local galaxies. Using nearly complete optical spectroscopy of emission lines, we made a classification based on the width of H$\alpha$ or the emission line ratios of [OIII]/H$\beta$ and [NII]/H$\alpha$. 
 We classified the AKARI sources into five types (Type 1 AGNs, Type 2 AGNs, LINERs, Composites, and SFs). 
 We then calculated the number density of Type 1 AGNs and Type 2 AGNs by integrating each LF. 
 The main results are as follows:

\begin{enumerate}
\item AKARI's efficiency at identifying AGNs is excellent, with 41\% of our 18 $\micron$ sources hosting active nuclei.
\item  The number density ratio of Type 2 to Type 1 AGNs obtained from the 18 $\micron$ LF is 1.73 $\pm$ 0.36. That value is larger than the results obtained from the optical ([OIII]) LF \citep{Haob}. The cause of this difference is probably the absorption of the [OIII] photons by the dust torus in Type 2 AGNs in particular. On the other hand, the results obtained from the MIR (12 $\micron$) and hard X-ray (15-55 keV) LFs are consistent with each other.
\item The fraction of Type 2s among all AGNs decreases with increasing 18 $\micron$ luminosity. Thus, the covering factor of the dust torus probably depends on the 18 $\micron$ luminosity. Our MIR results are also comparable to those reported using optical and X-ray observations.
\end{enumerate}
These results suggest that most of the AGNs in the local universe are obscured, but the obscuration (whether by a torus or another mechanism) decreases with MIR luminosity. 
 Furthermore, these obscured AGN are well-measured using MIR surveys, just as they are by complementary hard X-ray surveys.
 
Recently, the Wide-field Infrared Survey Explorer (WISE; \cite{Wright}) all-sky survey data was released. 
 WISE has performed a digital imaging survey of the entire sky in the 3.4, 4.6, 12, and 22 $\micron$ MIR bandpasses. 
 This has been used to construct AGN samples extending to substantial redshifts (e.g., \cite{Edelson}).
 The present study with AKARI provides us with an important local benchmark for AGN studies at high redshift.

\bigskip

 This study is based on observations made with AKARI. AKARI is a JAXA project with the participation of ESA.
The authors gratefully acknowledge the anonymous referee for a careful reading of the manuscript and very helpful comments.
 We also thank P.Gandhi (Durham University) and T.Goto (University of Copenhagen) for fruitful discussion.
 Funding for the SDSS and SDSS-II has been provided by the Alfred P. Sloan Foundation, the Participating Institutions, the National Science Foundation, the U.S. Department of Energy, the National Aeronautics and Space Administration, the Japanese Monbukagakusho, the Max Planck Society, and the Higher Education Funding Council for England. The SDSS Web Site is http://www.sdss.org/.
 The SDSS is managed by the Astrophysical Research Consortium for the Participating Institutions. The Participating Institutions are the American Museum of Natural History, Astrophysical Institute Potsdam, University of Basel, University of Cambridge, Case Western Reserve University, University of Chicago, Drexel University, Fermilab, the Institute for Advanced Study, the Japan Participation Group, Johns Hopkins University, the Joint Institute for Nuclear Astrophysics, the Kavli Institute for Particle Astrophysics and Cosmology, the Korean Scientist Group, the Chinese Academy of Sciences (LAMOST), Los Alamos National Laboratory, the Max-Planck-Institute for Astronomy (MPIA), the Max-Planck-Institute for Astrophysics (MPA), New Mexico State University, Ohio State University, University of Pittsburgh, University of Portsmouth, Princeton University, the United States Naval Observatory, and the University of Washington.
 This research is also made use of the NASA/IPAC Extragalactic Database (NED) which is operated by the Jet Propulsion Laboratory, California Institute of Technology, under contract with the National Aeronautics and Space Administration.
 This research has also made use of the SIMBAD database, operated at CDS, Strasbourg, France.


\end{document}